\documentclass[default,iicol]{sn-jnl}

\usepackage{graphicx}%
\usepackage{multirow}%
\usepackage{amsmath,amssymb,amsfonts}%
\usepackage{amsthm}%
\usepackage{mathrsfs}%
\usepackage[title]{appendix}%
\usepackage{xcolor}%
\usepackage{textcomp}%
\usepackage{manyfoot}%
\usepackage{booktabs}%
\usepackage{algorithm}%
\usepackage{algorithmicx}%
\usepackage{algpseudocode}%
\usepackage{listings}%
\usepackage{subfigure}
\usepackage{subcaption}
\usepackage[hang]{footmisc}
\renewcommand{\footnoterule}{%
  \kern -3pt
  \hrule width \dimexpr\textwidth/8\relax height 0.4pt
  \kern 2pt
}
\newcommand{\myblue}[1]{%
   \setcounter{footnote}{0}%
   \textcolor{blue}{#1};\ %
}

\theoremstyle{thmstyleone}%

\theoremstyle{thmstyletwo}%

\theoremstyle{thmstylethree}%

\begin{document}

\title[Article Title]{Contrastive Learning Method for Sequential Recommendation based on Multi-Intention Disentanglement}

\author*[1,2,3,4]{\fnm{Zeyu} \sur{Hu}}\email{81234hzy@gmail.com}

\author[1,2,3,4]{\fnm{Yuzhi} \sur{Xiao}}
\equalcont{Corresponding author. E-mail: \myblue{qh\_xiaoyuzhi@139.com}}

\author[1,2,3,4]{\fnm{Tao} \sur{Huang}}\email{762711327@qq.com}

\author[1,2,3,4]{\fnm{Xuanrong} \sur{Huo}}\email{2045519433@qq.com}

\affil[1]{\orgaddress{College of Computer, Qinghai Normal University, Xining, 810008, China}}

\affil[2]{\orgaddress{The State Key Laboratory of Tibetan Intelligent Information Processing and Application, Xining, 810008, China}}

\affil[3]{\orgaddress{Tibetan Information Processing and Machine Translation Key Laboratory of Qinghai Province, Xining, 810008, China}}

\affil[4]{\orgaddress{Key Laboratory of Tibetan Information Processing, Ministry of Education, Xining, 810008, China}}

\abstract{Sequential recommendation is one of the important branches of recommender system, aiming to achieve personalized recommended items for the future through the analysis and prediction of users' ordered historical interactive behaviors. However, along with the growth of the user volume and the increasingly rich behavioral information, how to understand and disentangle the user's interactive multi-intention effectively also poses challenges to behavior prediction and sequential recommendation.
In light of these challenges, we propose a \textbf{C}ontrastive \textbf{L}earning sequential recommendation method based on \textbf{M}ulti-\textbf{I}ntention \textbf{D}isentanglement (MIDCL).
In our work, intentions are recognized as dynamic and diverse, and user behaviors are often driven by current multi-intentions, which means that the model needs to not only mine the most relevant implicit intention for each user, but also impair the influence from irrelevant intentions.
Therefore, we choose Variational Auto-Encoder (VAE) to realize the disentanglement of users' multi-intentions. We propose two types of contrastive learning paradigms for finding the most relevant user's interactive intention, and maximizing the mutual information of positive sample pairs, respectively. Experimental results show that MIDCL not only has significant superiority over most existing baseline methods, but also brings a more interpretable case to the research about intention-based prediction and recommendation.
}


\maketitle

\section{Introduction}\label{sec1}

Nowadays, recommender system has gradually become an indispensable part of people's lives, meanwhile, the more realistic and logical sequential recommendation are becoming the key to provide personalized needs. As a popular research direction, sequential recommendation is not only for product recommendation, but also can be extended to other fields, such as intention understanding and prediction. The main reason is that the definition of sequential recommendation is not absolute, such as the task of recommending the next product to the user based on the browsing and purchasing records. But we can also consider it as a prediction task based on a series of consecutive historical interactive sequences of the user, and analyze the potential intention of users to generate the most probable interactions that will occur at the next moment, so that the method can be applied to real-world datasets in different scenarios.

\begin{figure}[htb]
  \centering
  \includegraphics[width=0.9\linewidth]{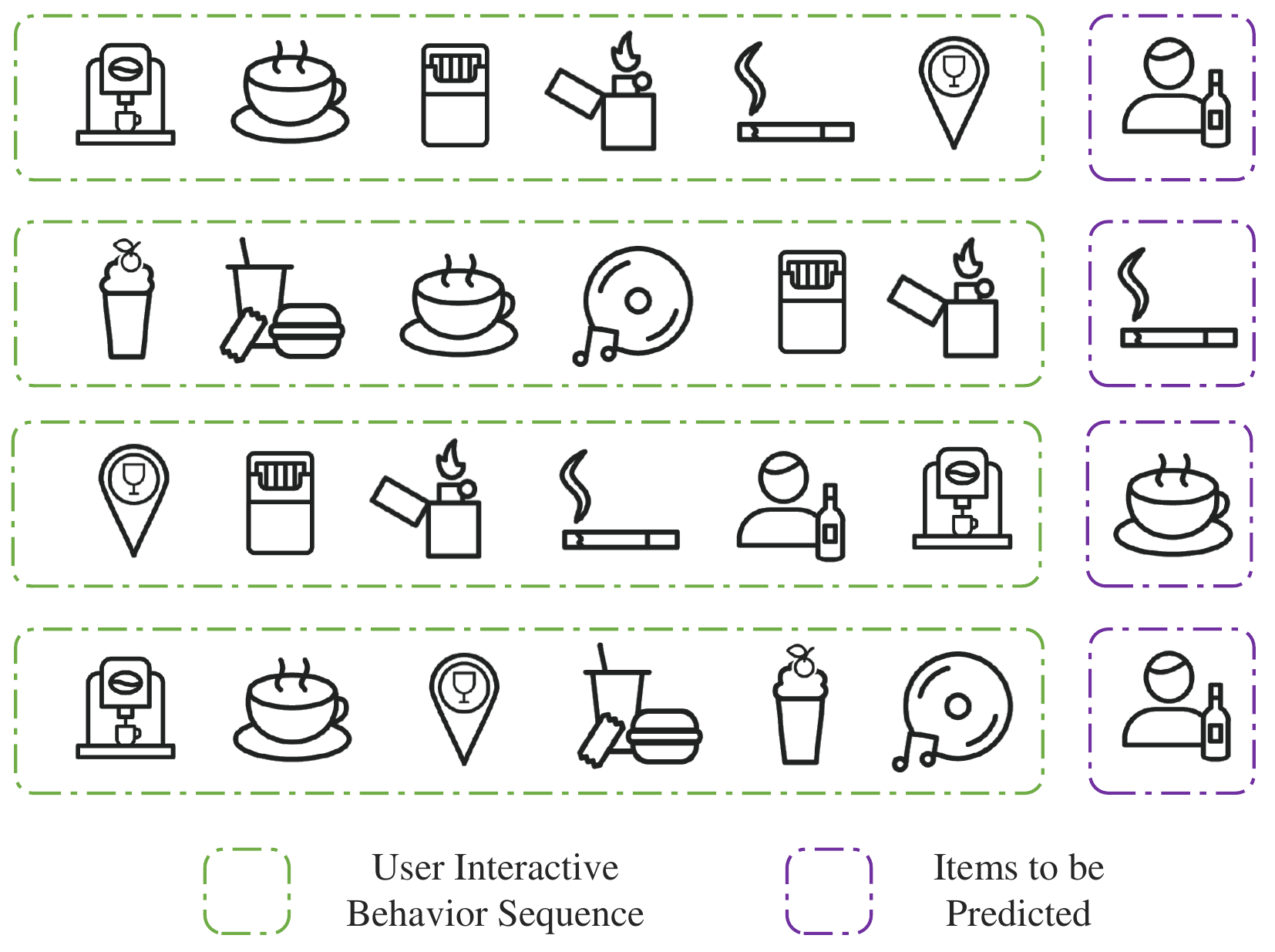}
  \caption{An example about historical behavioral sequences of different users, each item represents a kind of interactive behavior.}
  \label{fig1}
\end{figure}

The difference between prediction and recommendation is that recommendation usually does not recommend the interacted items to the user again, but the behavior sequence may contain a large number of repeated interactions, which leads to a wider scope of prediction than recommendation. Due to the real-time progression of interactive intention, we believe that the closer the interactive behavior is to the current moment, the more it can influence the prediction of the next moment. Fig.\ref{fig1} illustrates the sequence of four users' interactive behaviors, and it can be seen that there are a variety of coherent behaviors that contain certain fixed-pairing items. In particular, the most recent interacted item has the highest weight in the prediction task, but this is not absolute. As shown in this figure, it can be found that the first and last users have different interacted tail items, but the items to be predicted are the same. Therefore, it is reasonable to assume that although the two users have different behaviors at this point in time, their intentions are indeed close to each other, which reflects the importance of intention understanding and disentanglement.

The core task of sequential recommendation is to predict whether the user will interact with the candidate item at the next moment, and different scenarios will be handled in different ways. For example, in the graph-based recommendation scenario, we can regard it as a link prediction problem. Through the current research on sequential recommendation, it can be seen that most of the work focuses on the feature construction of the user and the item, but instead does not consider the historical behavior that contains rich potential information, and ignores the full mining of the user's interactive intention. Meanwhile, the input features of most models in Computer Vision (CV) and Natural Language Processing (NLP) are dense features, while the inputs of recommendation models are mostly sparse features such as id information, so how to design data enhancement methods for sparse features is the key to optimize the sequential recommendation task.

Most of the current researches have achieved surprising results in the prediction process as well as the recommendation accuracy, but we also want to build on these researches to further discover the user's interactive intention from the user's sequence, and endow the final recommendation results with a certain degree of interpretability as well as the generalization capability to face different dataset scenarios. Based on this, we propose a contrastive learning recommendation method based on the disentanglement of multi-intentions. First, the multiple entangled intentions of the user's current interactive behavior are disentangled and mapped to potential spaces through VAE, and the user's feature optimization is realized from the encoder to the decoder. Then, the two kinds of loss function is constructed based on the contrastive learning, and optimize feature representation according to Mutual Information Maximization (MIM) principles. Finally, use the learned user embedding and item embedding to compute dot product results to predict the probability of a user interacting with the item, thus enabling personalized recommendation.

The main contributions of this paper are as follows:

\begin{itemize}
    \item VAE-based disentanglement can discover and learn the latent intention of user's interactive behavior.
    \item Two contrastive learning paradigms are proposed that not only enhance the user's feature representation, but also weaken the negative impact of irrelevant intentions.
    \item MIDCL demonstrates good performance and generalization ability in experiments with four real scenario datasets, and also proves that intention disentanglement can compensate a certain degree of interpretability for the final recommendation results.
\end{itemize}

The work in this paper is structured as follows. In Sect.\ref{sec2}, we present relevant recent years studies. Subsequently, Sect.\ref{sec3} details the preliminaries about VAE and contrastive learning. In Sect.\ref{sec4}, we describe the proposed methodology in detail. Sect.\ref{sec5} verifies the validity of the proposed method through experiments and analysis. Finally, we discuss future work and provide conclusion for the paper in Sect.\ref{sec6}.

\section{Related work}\label{sec2}

\subsection{Sequential Recommendation}\label{sec2.1}
Unlike traditional recommendation based on collaborative filtering of users or items, sequential recommendation pays more attention to the interaction sequence of users' behavior history, focusing on mining users' real-time interests and interactive intentions. The earliest sequential recommendation started with statistical machine learning methods, based on Markov chain~\cite{PageRank} to predict the next clicked item through the user's historical sequence; combining the first-order Markov chain and matrix factorization~\cite{FPMC} not only focuses on the user's history, but also simulates the user's potential preferences; combining similarity computation in the Markov chain~\cite{Fossil} to a certain extent compensates for the shortcomings of the first two models.

With the rise of deep learning, researchers began to apply it to sequential recommendation because of its powerful ability to handle massive data and complex problems. For example, GRU4Rec~\cite{GRU4Rec}, which improves Recurrent Neural Network (RNN), alleviates the former problem of gradient vanishing or gradient explosion; Caser~\cite{Caser}, based on Convolution Neural Network (CNN), enhances feature associations between items by splicing the convolutional results in different directions; SASRec~\cite{SASRec} based on the self-attention mechanism~\cite{Transformer}, fully exploits the user's interest weight in the current interaction, and TiSASRec~\cite{TiSASRec} incorporates consideration of the time intervals between interactions based on ~\cite{SASRec}. BERT4Rec~\cite{BERT4Rec} consists of the bidirectional Transformer-based structure, which demonstrates its powerful processing and computing capabilities in the face of large-scale data.

Nevertheless, most of the existing algorithms for sequential recommendation lack the analysis of the user's dynamic interests and the interpretation of interaction intentions. 
In some studies, Graph Neural Network (GNN) has been employed to deal with heterogeneous networks~\cite{DHGAN, HySAGE, SURGE, MIDGN, GPR}, and some others try to discover latent user interests through contextual information ~\cite{CIPR, ASLI, DGAN} or interaction time-aware analysis ~\cite{TGSRec, TLSRec, GSMK},
but we hope to mine the potential multiple intentions of users from the interaction sequences themselves.
As a result, we build on the above and conduct a series of work that incorporates the intention during interaction behavior.

\subsection{Disentangled Representation Learning}\label{sec2.2}
In a sequential recommendation task, the user's choice will be influenced by contextual information, but also related to one's own interactive intention. Therefore, how to mine the potential intention is a current issue of great interest. Researchers believe that the user's interactive intentions are dynamic and entangled, so most of the research is devoted to decompose the diverse real-time interactive intentions through disentangled representation learning and flexibly apply them in prediction and recommendation.

VAE~\cite{VAE}, first proposed in 2013, is a generative network architecture based on Variational Bayes inference, designed for effective inference and learning in directed probabilistic models in the face of continuous latent variables with intractable posterior distributions and large datasets. VAE differ from Auto-Encoder in that the former introduce latent variables that are assumed to obey a prior distribution.

MacridVAE~\cite{MacridVAE} faces a shopping scenario with highly entangled user intentions and captures different dimensions of user preferences from macro and micro perspectives, respectively, with the help of VAE; MGNM~\cite{MGNM} simulates a recommendation scenario of a short-video and an e-commerce platform, where a loose sequence of items is combined into a single item by means of user perception to reconstruct loose item sequences into tight item-item interest graphs, and the introduction of sequence capsule networks in the iterative graph convolution inference process enhances the multi-interest extraction of users.

However, the results of directly applying most of the disentangled intentions in recommendation scenarios are often unsatisfactory due to the fact that the results not only rely on the prior distribution, but also suffer from the posterior collapse problem~\cite{ContrastVAE}. Therefore, after the disentanglement of user's intention, how to utilize these disentangled intentions is the core goal of our research, and mitigating the problem of representation degradation through contrastive learning is a widely recognized approach.

\subsection{Contrastive Self-Supervised Learning}\label{sec2.3}
Contrastive Learning, also known as Self-Supervised Learning or Unsupervised Learning, aiming to learn a representation or feature of the data by comparing the similarities and differences between different samples to learn the representation or features of the data. The key of contrastive learning in sequential recommendation-oriented tasks is how to construct suitable pairs of positive and negative samples.

CGCL~\cite{CGCL} provides high-quality embedding based on graph contrastive learning by constructing a structural neighborhood contrastive learning object, a candidate contrastive learning object and candidate structural neighborhood contrastive learning object nodes. In addition, contrast learning is usually combined with intention disentanglement. Paper~\cite{seq2} proposes a novel seq2seq training strategy to disentangle multiple user's intentions and cluster the intentions based on prototype distances to enable prediction of future intentions; IOCRec~\cite{IOCRec} fuses local and global intentions to unify sequential patterns and intention-level self-supervised signals, and demonstrates a good denoising performance that alleviate the impact of different user's intentoins on the prediction results; ICLRec~\cite{ICLRec} uses K-means clustering to represent the distribution of users' potential intentions and maximizes the consistency between sequential views and their corresponding intentions based on the idea of expectation-maximization, from which the loss function is constructed and trained alternatively; ICSRec~\cite{ICSRec} uses cross-sequence contrastive learning to model potential user purchase intention by splitting the user interaction sequence into multiple subsequences and encoding them, assumes that subsequences with the same target item have the same user intent, and uses intention-based contrastive learning to pull together subsequences with the same target item; DCCF~\cite{DCCF} proposes a framework about disentangled contrastive with collaborative filtering, in order to enable intention disentanglement with self-supervised augmentation in adaptive way.

S\textsuperscript{3}-Rec~\cite{S3-Rec}, as a sequential recommendation model based on a self-attentive neural architecture, proposes four classes to learn the correlation between attributes, items, subsequences and sequences respectively by using the MIM principle; CL4SRec~\cite{CL4SRec} optimizes the user feature representation based on maximizing the consistency between different augmented views of the same user interaction sequence in the latent space, and similarly proposes three training methods based on contrastive learning by augmenting the data i.e. cropping, masking, and reordering, respectively. Both of the above models provide valuable reference solutions for data optimization for future contrast learning tasks, and CoSeRec~\cite{CoSeRec} proposes two additional informative enhancement operators on top of~\cite{CL4SRec}, which are substituting and inserting.

Unlike above methods, in our work, in addition to contrastive learning of sequences, we propose a new paradigm for contrastive learning of intentions, which is dedicated to weakening the influence of irrelevant intentions on user's behavior reasoning.

\section{Preliminaries}\label{sec3}

\subsection{Problem Definition}\label{sec3.1}
Let $\mathcal{U}$ denote the set of users $\{u_1,u_2,\dots,u_{|\mathcal{U}|}\}$,
$\mathcal{I}$ denote the set of items $\{i_1,i_2,\dots,i_{|\mathcal{I}|}\}$.
Define $H^u = [H^u_1, H^u_2, \dots, H^u_{|H^u|}]$ as the sequences of interaction items of user $u$ sorted by timestamp to moment $T$, where $H^u_T$ donates the item that $u$ has interacted with most recently.
For each user, $k$ is the number of disentangled intentions.

The goal of our work can be formulated as predicting the item $i_u^*$ that the current user $u$ is most likely to interact with at the next moment, after the disentanglement of $u$'s most relevant intention $e^{msi}_i$:
\begin{equation}
\label{eq:predict}
i^*_u = \underset{e^i,i\in \mathcal{I}}{\mathrm{arg\,max}}~
P(H^u_{T+1} = e^i \vert e^{msi}_i,H^u_T).
\end{equation}

\subsection{Intention Disentanglement}\label{sec3.2}
The essence of VAE is a generative model, whose structure consists of an encoder and a decoder. VAE is an extension from the self-encoder, which works on the idea of obtaining an ideal distribution of the sample data through the encoder based on a portion of the real samples, then reconstructing new samples in accordance with this distribution through the decoder, and the more ideal the distribution, the closer the new samples can make these new samples to the original real samples. VAE, on the other hand, makes further variance processing on the basis of the self-encoder, it allows the model to learn the potential distribution of the sample data and at the same time also learns the mean and variance of the sample distribution.

Assume a batch of data samples $X = \{x_1, \dots x_n\}$, by obtaining the distribution $p(X)$ of $X$, and then sampling according to $p(X)$, a new sample $Z$ can be obtained based on the $X$ distribution. Derive this distribution as the posterior distribution of the hidden variables:
\begin{equation}
\label{eq:p(x)}
p(X) = \sum_Z p(X \vert Z) p(Z).
\end{equation}

\noindent The $Z$ in Eq.(2) is expected to satisfy the standard normal distribution, which means $p(Z)=\mathcal{N}(0,1)$, and such a distribution can help the decoder generate more similar samples. The normal distribution is affected by the mean $\mu$ and variance $\sigma^2$, and VAE can also use a neural network to help it fit the vector representation of $\mu$ and $\sigma^2$. For a real sample $X_k$, set up a distribution $p(Z \vert X_k)$ exclusively belonging to $X_k$, train the $\mu$ and $\sigma^2$ of this distribution with encoder by approaching to the normal distribution through a neural network $X=g(Z)$, and then use decoder to reconstruct $Z_k$ sampled from the distribution $p(Z \vert X_ k)$ to $X_k$. Specifically, the neural network $X$ contains two encoders $f_1(X_k)$ and $f_2 (X_k)$, which are used to train $\mu_k$ and $\log{\sigma^2_k}$,
respectively, and reconstructs the samples by a generator
$\hat{X}_k = g(Z_k)$,
constructing a loss function with a mean-square error as 
$\mathcal{D}(\hat{X}_k,X_k)^2$
and the sum of Kullback-Leibler Divergence (KLD) as 
$KL\bigl( \mathcal{N}(\mu,\sigma^2) \Vert \mathcal{N}(0,1) \bigr)$,
the VAE turns the sampling from $\mathcal{N}(\mu,\sigma^2)$ into sampling from
$\mathcal{N}(0,1)$,
and then obtains the sampling from $\mathcal{N}(\mu,\sigma^2)$ by a parameter transformation:
\begin{equation}
\label{eq:VAE_Loss}
VAE_{Loss} = \mathcal{D}(\hat{X}_k,X_k)^2 + KL\bigl( \mathcal{N}(\mu,\sigma^2) \Vert \mathcal{N}(0,1) \bigr).
\end{equation}

\noindent The `Variational' in VAE means that the derivation process uses the variational method to find the variational lower bound of the KLD generalized extremum, which is essentially the introduction of Gaussian noise into the encoder on the basis of the self-encoder, so that the decoder can be robust to the noise to improve the generalization ability. The KLD is equivalent to a regular term of the encoder, which aims to keep $p(Z)$ as close as possible to the standard normal distribution. Conversely, the encoder is also capable of dynamically adjusting the intensity of the noise, thus achieving a GAN-like adversarial training process. In order to satisfy the individual needs of users, in our work, the VAE is allowed to learn the mean and log-variance independently for each user in each batch, and to make the distribution of each pair of mean and log-variance converge to the normal distribution during the training process.

\subsection{Contrastive Learning}\label{sec3.3}
Contrastive learning originated in 2014 when Kateria Fragkiadaki et al. questioned in the face of supervised pre-training of ImageNet~\cite{R-CNN}, whether semantic supervision was really needed to learn good information representations. Five years later, the MoCo model based on the contrast learning paradigm proposed by He et al.~\cite{MoCo} outperformed the mainstream supervised learning methods on the task of Pascal VOC target detection of the time. Contrastive learning is centered on gradually narrowing the spacing of similar samples and expanding the distance of dissimilar samples by comparing the differences between two or more samples, and has been widely used because it does not require artificial labels. Let a data sample $x$, contrastive learning aims to learn an encoder $f(x)$ such that:
\begin{equation}
\label{eq:CL_1}
Score\bigl( f(x),f(x^+) \bigr) \gg Score\bigl( f(x),f(x^-) \bigr),
\end{equation}

\noindent where $x^+$ denotes a positive sample that is similar to $x$, and $x^-$ denotes a negative sample that is not similar to $x$. $Score(\cdot,\cdot)$ is a measure of the similarity between two features, typically a function of Euclidean distance, cosine similarity, etc. Here $x$ is commonly referred to as the anchor point, and aiming to optimize $x$, a classifier is typically constructed to assign high scores to positive samples and low scores to negative samples:
\begin{equation}
\label{eq:CL_2}
\begin{split}
& \mathcal{L}_N = -\mathbb{E}_X 
 \\
& [\log \frac{\exp \bigl( f(x)^\top f(x^+) \bigr)}{\exp \bigl( f(x)^\top f(x^+) \bigr) + \sum_{j=1}^{N-1} \exp \bigl( f(x)^\top f(x_j) \bigr)}],
\end{split}
\end{equation}

\noindent and
\begin{equation}
\label{eq:CL_3}
Score\bigl( f(x),f(x^+) \bigr) = f(x)^\top f(x^+),
\end{equation}

\noindent where the numerator is a positive sample, the denominator includes one positive sample and N-1 negative samples, the similarity metric is the dot product. In contrastive learning, this loss function for evaluating positive and negative samples with anchors is called InfoNCE~\cite{InfoNCE} (Information Noise-Contrastive Estimation).

In contrastive learning, mutual information~\cite{MI} is used to indicate the degree of association of an anchor point with a positive sample, so the process of minimizing the InfoNCE loss is also seen as maximizing the mutual information lower bound of $f(x)$ and $f(x^+)$, and VAE also serves the same purpose of constraining the Evidence Lower Bound while minimizing the prior distribution role. Specifically, the correlation between samples $X$ and $Z$ is expressed in terms of mutual information:
\begin{equation}
\label{eq:CL_4}
\begin{split}
I(X,Z) & = \iint {p(z \vert x) \tilde{p}(x) \log \frac{p(z \vert x) \tilde{p}(x)}{p(z) \tilde{p}(x)}} dxdz
 \\
& = KL(p(z \vert x) \tilde{p}(x) \Vert p(z) \tilde{p}(x)),
\end{split}
\end{equation}

\noindent where $\tilde{p}(x)$ represents the distribution of the original data, $p(z \vert x) \tilde{p}(x)$ describes the joint distribution of $x$ and $z$, and $p(z)$ denotes the distribution of the entire $Z$ after given $p(z|x)$:
\begin{equation}
\label{eq:CL_5}
p(z) = \int p(z \vert x) \tilde{p}(x) dx.
\end{equation}

\noindent VAE wants to maximize the mutual information $I(X,Z)$, i.e., to widen the distance between $p(z \vert x) \tilde{p}(x)$ and $p(z) \tilde{p}(x)$:
\begin{equation}
\label{eq:CL_6}
p(z \vert x) = \underset{p(z \vert x)}{\mathrm{arg\,max}} I(X,Z).
\end{equation}

\noindent Let $q(z)$ be the prior distribution of the standard normal distribution, and write the loss of KLD in Eq.(3) as the sum of two components:
\begin{equation}
\label{eq:CL_7}
\begin{split}
& KL \bigl( {p(z) \Vert q(z)} \bigr) = KL \bigl( {\tilde{p}(x) p(z \vert x) \Vert q(z)q(x \vert z)} \bigr)
 \\
& = \iint{\tilde{p}(x) p(z \vert x) \log{\frac{\tilde{p}(x) p(z \vert x)}{q(x \vert z)q(z)}}} dzdx
  \\
& = \iint{\tilde{p}(x) p(z \vert x) \log{\frac{p(z \vert x)}{q(z)}}} dzdx
 \\
& - \iint{\tilde{p}(x) p(z \vert x) \log{\frac{q(x \vert z)}{\tilde{p}(x)}}} dzdx,
\end{split}
\end{equation}

\noindent where the first term is the KLD of the prior distribution and the second term is the mutual information of $x$ and $z$. The purpose of VAE is to maximize the mutual information while minimizing the prior distribution.

The idea of contrastive learning was first applied in the field of CV~\cite{DIM, AMDIM}. Let the global features of an image be the anchor point, positive samples are local features from the same image, and negative samples are local features from different images. With the research in recent years, contrastive learning and VAE have also been introduced into the domain of sequential recommendation. In our work, the method will be proposed to disentangle the potential user's intention distribution based on VAE, and we also propose two new paradigms combined with contrastive learning, aiming to help the model to get a more accurate user sequence and intention representation.

\section{Methodology}\label{sec4}

\subsection{Model Architecture}\label{sec4.1}

\begin{figure*}[htb]
  \centering
  \includegraphics[width=0.95\textwidth]{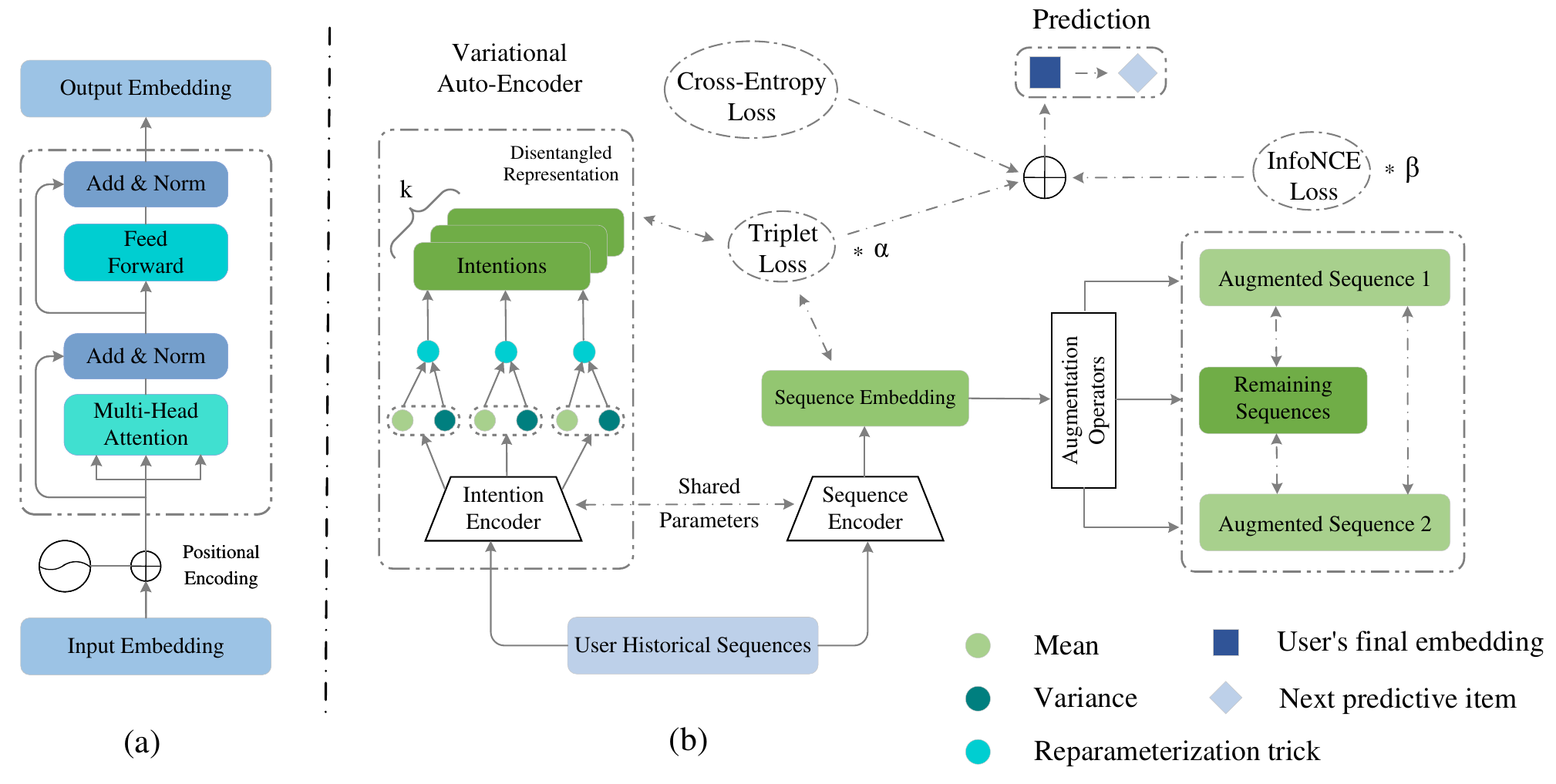}
  \caption{Overview of MDICL. (a) shows the basic structure of Transformer encoder. (b) describes the framework and process of MIDCL, which includes the intention disentanglement by VAE and contrastive learning. The detailed illustration of VAE is in Sect.\ref{sec4.3}, and five kinds of augmentation operators is recorded in Sect.\ref{sec4.4}. Multi-task training integrates adaptive results from three types of loss functions for the final prediction.}
  \label{fig2}
\end{figure*}

\indent The model architecture of MIDCL is shown in Fig.\ref{fig2}. Where (a) is the encoder part of the Transformer, the position encoding is changed from the original sine-cosine encoding to the learnable position encoding.
(b) is the structure of intention disentanglement and the construction process of the loss function, the input is the user's historical interaction sequence, and then two Transformers with shared parameters encode the user's intention vectors and sequence vectors, respectively. The distribution of the intention is learnt by the VAE, and the user's sequence encoding is is optimized using the attention mechanism.
In the part of contrastive learning, we propose two kinds of paradigms, one of them is intention contrastive learning, the anchor point is the user's optimized sequence encoding, the positive samples are the most similar intentions, the negative samples are the least similar intentions, and the loss function is the Triplet loss, which brings the most similar intentions closer to the most similar intentions and as far as possible away from the least similar intentions by minimizing the loss of the triplet.

The second one is sequence contrastive learning, the anchor point is the interaction sequence of one of the $N$ users, after the random operator augmentation, the positive samples are the 2 augmented sequences of the current user, and the negative samples are the $2(N-1)$ augmented sequences of the rest of users, the positive and negative samples go through the same encoder to generate vectors, and then optimize the features according to the InfoNCE loss function, which is designed to maximize the mutual information of the positive samples from the same sequence, and minimize the mutual information of the negative samples from different sequences.

\subsection{Embedded Layer}\label{sec4.2}
In our work, we uniformly use Transformer as encoder, compared to CNN, Transformer's sensory field is more suitable for sequence data, and compared to RNN, Transformer is more adept at learning the item impact of each item in the sequence, and does not suffer from the problem of disappearing or exploding gradient in the training process. For the interaction sequence $[H_1^u,H_2^u,\dots,H_T^u]$ of user $u$, generate an embedding representation for the items at each position:
\begin{equation}
\label{eq:EL_1}
e_i^u = \mathrm{Trm}(H_T^u + p_T),
\end{equation}

\noindent where $\mathrm{Trm}(\cdot)$ is the Transformer encoder and $p_T$ is the learnable positional encoding of $H_T^u$ at moment $T$. Transformer's parallel computation of each item in the sequence leaves it with no way to obtain the absolute position of the item in the sequence, which is a fatal problem in the field of NLP because the position of a single word may change the meaning of the whole sentence, so the introduction of positional encoding with sine-cosine computation enables the position of each word or item in the sequence, that is, the effect on the whole sequence, to be represented.
Compared to the sine-cosine function coding in paper~\cite{Transformer}, many studies have suggested that learnable absolute positional coding can bring better performance, and that positional coding learned through neural networks has stronger expressive power, which can be more adaptive to varying sequence lengths, possesses higher scalability, and is better suited for end-to-end training tasks because learnable positional coding is learned together as an overall parameter of the model, rather than as a stand-alone process.

Doing so, however, ignores the relative position of each item in the sequence. Since for the task of sequential recommendation, the user's interactive intention changes over time, and the last item in the sequence, i.e., the intention of the last interaction behavior, has the greatest impact on predicting the interaction behavior at the next moment, the model needs to take into account the weighting factor of each item for the last interaction. Although the models in paper~\cite{BLEU, RoPE, LLaMA} take the relative position or rotational position coding of each item (relative position coding implemented by the absolute position coding approach), it is also necessary to consider the effect of the item itself on the overall sequence with the final intention. Therefore, in the study in this paper, the user's sequence characteristics are optimized by adopting an attention mechanism for the first $T-1$ items versus the last interacted item.

The user's historical sequence $H^u$, sorted by interaction timestamp, is used as input to the Transformer encoder, which encodes an embedded representation $[e_1^u,\dots,e_{T-1}^u,e_T^u]$ for each position. This result incorporates the multi-head self-attention mechanism of the $h$-heads, set to be encoded by each head:
\begin{equation}
\label{eq:EL_2}
Head_{i \in h} = \mathrm{SelfAtt}(Q_{H^u}, K_{H^u}, V_{H^u}),
\end{equation}

\noindent where $\mathrm{SelfAtt}(\cdot)$ is the self-attention mechanism, and $Q_{H^u}$, $K_{H^u}$, $V_{H^u}$ are the query, key, and value derived from $H^u$ through the three parameter matrices $W_i^q$, $W_i^k$, and $W_i^v$, respectively. Each head focuses on a different level of behavioral information, and concatenates the results of $h$ heads to get the multi-head representation
$[e_{mh_1}^u, \dots, e_{mh_{(T-1})}^u, e_{mh_T}^u]$:
\begin{equation}
\label{eq:EL_3}
e_{mh}^u = \mathrm{Dropout}(Head_1 \Vert Head_2 \Vert \dots \Vert Head_h).
\end{equation}

\noindent Based on residual networks~\cite{ResNet}, combine the Add\&Norm and Feed Forward modules in Transformer encoder to get the encoding for each position:
\begin{equation}
\label{eq:EL_4}
\mathrm{AN}(e_i^u) = \mathrm{LayerNorm}(e_{mh}^u + e_i^u),
\end{equation}
\begin{equation}
\label{eq:EL_5}
\mathrm{FFN}(e_i^u) = \mathrm{ReLU} \bigl( {\mathrm{AN}(e_i^u) W_1 + b_1} \bigr) W_2 + b_2.
\end{equation}

\noindent Since the goal of the model is to predict the most likely interaction item at the moment $T+1$, and the user's interactive intention is changing in any order, $e_{mh_T}^u$ has a greater impact on the final prediction result compared to the first $T-1$ items. Therefore, based on the attention mechanism, the attention score of each item with $e_{mh_T}^u$ is calculated and normalized:
\begin{equation}
\label{eq:EL_6}
\mathrm{AttScore}_{i,T} = \mathrm{Softmax} \bigl( {\mathrm{Dot}(e_{mh_i}^u, e_{mh_T}^u)} \bigr),
\end{equation}
\begin{equation}
\label{eq:EL_7}
\mathrm{Att}(e_{mh_T}^u) = \sum_{i=1}^{T-1} \mathrm{AttScore}_{i,T} \cdot e_{mh_i}^u,
\end{equation}

\noindent where $\mathrm{Dot}(\cdot,\cdot)$ is generally the dot product or scaled dot product, because the $T$-th term has the highest attention weight with itself, and in order to avoid the influence of itself on the other items, $e_{mh_T}^u$ is added after the weighted summation:
\begin{equation}
\label{eq:EL_8}
e^{trm} = e_{mh_T}^u + \mathrm{Att}(e_{mh_T}^u).
\end{equation}

\noindent This yields the final sequence representation $e^{trm}$ for $u$. Compares to the traditional direct splicing, this feature both includes the $T$-moment items and absorbs the effect of the weights of the rest of the $T-1$ items on the last item, with higher weights representing the greater importance of the interaction item at that moment for predicting the $T+1$ moment.

\subsection{Intention Disentanglement Layer}\label{sec4.3}
Our work argues that the interactive intentions in user behavior are complex, varied and highly disentangled, but most current research has neglected the effective parsing of user intentions. After obtaining the user's sequence representation, MIDCL learns the user's latent distribution representation based on VAE and generates a $k$-dimensional latent vector
$e^{int} \in \mathbb{R}^{k \times d}$
via an encoder, with $k$ denoting the number of disentangled intentions. VAE is divided into two encoders and a decoder, the encoders map the samples into the latent space, learning the mean and logarithmic variance of the sample distributions, respectively, and this latent space is usually constrained to be close to the standard normal distribution by the KLD and other loss functions. The decoder samples from the latent space and maps back into the data space based on the learned parameters, reconstructing new samples with similar distributions to the original samples. MIDCL discards the decoder portion of the VAE and samples the latent intention distributions of the user sequences using the encoder only, and the learning process relies on the reparameterization technique.

Most VAE encoders and decoders are linear layers, in order to enable them to learn deeper sample distributions, MIDCL adopts Transformer as the feature representation of encoder-generated user-sequence samples, and uses two linear layers to learn the representations of the sample distributions with the mean $\mu$ and the logarithmic variance $\log(\sigma^2)$ respectively:
\begin{equation}
\label{eq:EL_9}
e^{mu} = \mathrm{Linear}_1(e^{trm}),
e^{lv} = \mathrm{Linear}_2(e^{trm}).
\end{equation}

\noindent In order to make the encoder output intention representation can be in the same feature space with the sequence representation so as to achieve contrastive learning, compared with the direct learning of the variance, the logarithmic variance avoids the extreme value problem that will occur in the variance, which has better stability, and also makes the optimization process of the model smoother due to the logarithmic transformation to reduce the sharp curvature of the parameter space. Where $\mu$ controls the center of the data distribution and $\log(\sigma^2)$ constrains the dispersion of the data points, both of which together determine the latent intention in the space. The reparameterization trick is reflected in the sampling process:
\begin{equation}
\label{eq:EL_10}
\sigma = \sqrt{\exp \bigl( {\log(\sigma^2)} \bigr)},
\end{equation}
\begin{equation}
\label{eq:EL_11}
z = \mu + \epsilon \cdot \sigma = \mu + \epsilon \cdot {\sqrt{\exp \bigl( {\log(\sigma^2)} \bigr)}},
\end{equation}
\begin{equation}
\label{eq:EL_12}
e^{int} = e^{mu} + \epsilon \cdot \exp(\frac{e^{lv}}{2}),
\end{equation}

\noindent where Eq.(22) is the feature calculation of Eq.(21), $\sigma$ is the standard deviation, and $\epsilon$ is the Gaussian noise shaped like $\sigma$ sampled from the standard normal distribution $\mathcal{N}(0,1)$. The randomly sampled Gaussian noise allows the VAE to learn that $\mu$ and $\log(\sigma^2)$ bring the original sample distribution close to $\mathcal{N}(0,1)$ as a prior distribution. $\epsilon \cdot \sigma$ allows the random sampling process to be differentiable, which allows the model to be optimized by the training methods of backpropagation and gradient descent, the computation which essentially scales $\epsilon$ so that it has the encoder output's variance and also ensures the differentiability of the sampling step.

The VAE in MIDCL retains only the encoder and discards the reconstruction works of the decoder, so the part of the loss function abandons the reconstruction loss, and only the KLD is used as a measure of the difference between the distribution of the latent variable and the sampling source $\mathcal{N}(0,1)$ of $\epsilon$:
\begin{equation}
\label{eq:EL_13}
\mathcal{L}_{KL} = -\frac{1}{2} \sum_{i=1}^k
\bigl( 1+\log(\sigma_i^2)-\mu_i^2-\sigma_i^2 \bigr).
\end{equation}

\noindent The discarded reconstruction loss will remain in the overall optimization content of the model as part of the prediction outcome loss and intention contrastive learning. The process of minimizing the loss function enables the $k$-dimensional latent intentions learned and disentangled by VAE to be more independent and less entangled for the next step of contrastive learning.

\subsection{Contrastive Learning Layer}\label{sec4.4}
\subsubsection{\textbf{Intention-based Contrastive Learning}}
The contrastive learning of intentions is optimized using the Triplet loss~\cite{TRI} function to optimize the learning process of VAE. The anchor point is the sequence feature $e_s^u$ of $u$, the positive sample is the intention $I_{pos}^u$ with the highest similarity to $e_s^u$, and the negative sample is the intention $I_{neg}^u$ with the lowest similarity to $e_s^u$:
\begin{equation}
\label{eq:ICL_1}
\begin{split}
I_{pos}^u = \mathrm{arg\,max} \bigl( sim(e_s^u,e^{int}) \bigr),
 \\
I_{neg}^u = \mathrm{arg\,min} \bigl( sim(e_s^u,e^{int}) \bigr),
\end{split}
\end{equation}

\noindent where the intentions of positive and negative samples are derived from the similarity matrix of sequence features and intention features, and are extracted by index, the found positive and negative samples are used to construct the loss function:
\begin{equation}
\label{eq:ICL_2}
\mathcal{L}_{TRI} = \sum_{u=1}^N \bigl[\|{e_s^u-I_{pos}^u}\|_2^2
- \|{e_s^u-I_{neg}^u}\|_2^2 + \xi \bigr]_+,
\end{equation}

\noindent where $\xi$ is the hyperparameter used to constrain the minimum distance between positive and negative samples. Eq.(25) expects to bring the anchor point as close as possible to the distance between positive samples and as far away from the distance between negative samples, as a way of mitigating the noise-negative optimization that invalid intentions associated with the current user may bring to the final representation of the intention. Since the spacing between positive and negative samples may be less than $\xi$, the result of Eq.(25) always takes a non-negative value. Combining this with the KLD from above yields an overall loss in VAE:
\begin{equation}
\label{eq:ICL_3}
\mathcal{L}_{VAE} = \mathcal{L}_{KL} + \mathcal{L}_{TRI}.
\end{equation}

\subsubsection{\textbf{Sequence-based Contrastive Learning}}
Reference~\cite{CL4SRec, CoSeRec}, three stochastic enhancement operators and two informative enhancement operators are introduced into the augmented operator called $AO$ to provide positive and negative samples for contrastive learning. In order to ensure that the tail items have an effect on the sequence coding, we will not make any enhancements to the last item.

\noindent (1) \textbf{Stochastic Augmentation Operator}.

\begin{itemize}
    \item \textbf{Crop}:
    Remove a portion of elements in the input sequence, typically a continuous segment of the sequence. By cropping, the model needs to learn to make inferences in the missing part, thus improving the understanding of the local structure of the sequence. The length of the cropping is controlled by the parameter $\gamma_c$ and the position of the cropping is determined by the parameter $c$:
    \begin{equation}
    \label{eq:SCL_1}
    \begin{split}
    & L_c = \lfloor {\gamma_c \times |H^u|} \rfloor,
     \\
    & H_{Crop}^u = [H_c^u, \dots, H_{c+1}^u, \dots, H_{c+L_c-1}^u],
     \\
    & AO_{Crop}(H^u) = H^u - H_{Crop}^u.
    \end{split}
    \end{equation}

    \item \textbf{Mask}:
    Some elements in the input sequence are masked out, which are often replaced by special markers (e.g. $[MASK]$), similar to the pre-training approach of the Cloze Task in BERT. The number of masks is controlled by the parameter $\gamma_m$, and the positions of masks are the values of $L_m$ elements randomly selected from the sequence $[1,2,\dots,|H^u|]$ as the index of the markers of $[MASK]$:
    \begin{equation}
    \label{eq:SCL_2}
    \begin{split}
    & Len(H_{Mask}^u) = L_m = \lfloor {\gamma_m \times |H^u|} \rfloor,
     \\
    & AO_{Mask}(H^u) = \bigl[ H_1^u, \dots, [MASK], \dots, H_{|H^u|}^u \bigr].
    \end{split}
    \end{equation}

    \item \textbf{Reorder}:
    Randomly change the order of elements in a sequence. The reordering allows the model to learn to deal with different arrangements of elements and increase its ability to understand the global structure of the sequence. The length of the reordering is controlled by the parameter $\gamma_r$, the positions of the elements of the selected sequence are randomly disrupted, and the starting position is determined by the parameter $r$:
    \begin{equation}
    \label{eq:SCL_3}
    \begin{split}
    & L_r = \lfloor {\gamma_r \times |H^u|} \rfloor,
     \\
    & H_{Reorder}^u = \mathrm{Random}(H_r^u, \dots, H_{r+1}^u, \dots, H_{r+L_r-1}^u),
     \\
    & AO_{Reorder}(H^u) =\bigl[ H_1^u, \dots, H_{Reorder}^u, \dots, H_{|H^u|}^u \bigr].
    \end{split}
    \end{equation}
\end{itemize}

\noindent (2) \textbf{Informative Augmentation Operator}.

\begin{itemize}
    \item \textbf{Insert}:
    Inserting similar elements into the input sequence, i.e., the most similar items of the item, is generally measured by cosine similarity. The insertion method can give the model to adapt to longer sequences and feel richer contextual information. The number of insertions is controlled by the parameter $\gamma_i$, and the location of the insertions ${Sample}_i$ is randomly sampled:
    \begin{equation}
    \label{eq:SCL_4}
    \begin{split}
    & Len({Sample}_i) = L_i = \lfloor {\gamma_i \times |H^u|} \rfloor,
     \\
    & {Sample}_i = \{ I_1, I_2, \dots, I_{L_i} \vert I_i \in |H^u| - 1 \},
     \\
    & AO_{Insert}(H^u) 
     \\
    & =\bigl[ H_1^u, \dots, \widehat{H_I^u}, H_I^u, \dots, H_{|H^u|}^u \bigr], I \in {Sample}_i.
    \end{split}
    \end{equation}

    \item \textbf{Substitute}:
    Replacing selected items with the most similar items in the input sequence. Substitution helps the model to adapt to different elements and improves robustness to changes in the sequence as a whole or in the items. The number of substitutions is controlled by the parameter $\gamma_s$, and the location of the substitutions is randomly sampled ${Sample}_s$:
    \begin{equation}
    \label{eq:SCL_5}
    \begin{split}
    & Len({Sample}_s) = L_s = \lfloor {\gamma_s \times |H^u|} \rfloor,
     \\
    & {Sample}_s = \{ S_1, S_2, \dots, S_{L_s} \vert S_i \in |H^u| - 1 \},
     \\
    & AO_{Substitute}(H^u) 
     \\
    & =\bigl[ H_1^u, \dots, \widehat{H_S^u}, \dots, H_{|H^u|}^u \bigr], S \in {Sample}_s.
    \end{split}
    \end{equation}
\end{itemize}

\noindent Let $\mathcal{A}$ be a randomly selected augmentation operator from $AO$, and each batch of sequences ${\{H^u\}}_{u=1}^N$ from $N$ users is input into the sequence contrastive learning module, and each time two kinds of $\mathcal{A}$ are selected to generate the augmented sequences:
\begin{equation}
\label{eq:SCL_6}
\begin{split}
\mathcal{A}_1(\{H^u\}) = [H_{a_1}^1, \dots, H_{a_1}^u, \dots, H_{a_1}^N],
 \\
\mathcal{A}_2(\{H^u\}) = [H_{a_2}^1, \dots, H_{a_2}^u, \dots, H_{a_2}^N].
\end{split}
\end{equation}

\noindent For $u$, $2N$ augmented sequences are obtained, only $2$ of which belong to itself and the remaining $2(N-1)$ are not. The sequences belonging to $u$ itself are extracted as positive samples and the other sequences as negative samples, which are input to the encoder given by Sect.\ref{sec4.2} to generate the augmented sequence features:
\begin{equation}
\label{eq:SCL_7}
\begin{split}
& e_{pos_1}^u = \mathrm{Encoder}(H_{a_1}^u),
e_{pos_2}^u = \mathrm{Encoder}(H_{a_2}^u),
 \\
& e_{neg}^u = \mathrm{Encoder}(H_{neg}^u),
\end{split}
\end{equation}

\noindent where $e_{pos_1}^u$ and $e_{pos_2}^u$ are positive sample features and $e_{neg}^u$ is a negative sample feature. Construct InfoNCE loss function:
\begin{equation}
\label{eq:SCL_8}
\mathcal{L}_{SCL} = \mathcal{L}_{SCL}(e_{pos_1}^u,e_{pos_2}^u)
+ \mathcal{L}_{SCL}(e_{pos_2}^u,e_{pos_1}^u),
\end{equation}

\noindent and
\begin{equation}
\label{eq:SCL_9}
\begin{split}
& \mathcal{L}_{SCL}(e_{pos_1}^u,e_{pos_2}^u) = 
 \\
& - \log \frac
{\exp \bigl( sim(e_{pos_1}^u,e_{pos_2}^u) / \tau \bigr)}
{\sum_{neg} \exp \bigl( sim(e_{pos_1}^u,e_{neg}^u) / \tau \bigr)},
\end{split}
\end{equation}

\noindent where $sim(\cdot,\cdot)$ is the dot product similarity, $\tau$ is the temperature coefficient, used to control the sharpness of the sample similarity. The smaller the value the more the model can focus on the distinction between positive and negative samples, the larger the value makes the probability distribution the smoother, so as to weaken the model's sensitivity to the noise, and improve the robustness. Eq.(35) wants to expand the similarity between $H_{a_1}^u$ and $H_{a_2}^u$, and reduce the similarity between $H_{a_1}^u$ and $H_{neg}^u$, replacing the positive samples is in the same way. Eq.(34) is the sum of the losses of the two positive samples as the overall loss function of the contrastive learning layer.

\subsection{Multi-Task Training}\label{sec4.5}
The user's sequence features and intention features both contain potential information about the user, and the concatenation of two features as the final representation of the user's input to the prediction layer, which calculates the dot product with the user's to-be-predicted items and returns the interaction probability:
\begin{equation}
\label{eq:MT_1}
IP_i^u = \sigma \bigl( \mathrm{MLP}(e_s^u \Vert I_{pos}^u) \cdot e_i^u \bigr)
, e_i^u \in NI^u,
\end{equation}

\noindent where $\sigma$ is the $\mathrm{Sigmoid}$ function and $NI^u$ is the set of $u$'s to-be-predicted items. For the predicted loss of the interaction item at the next moment, the Cross-Entropy loss function is used:
\begin{equation}
\label{eq:MT_2}
\mathcal{L}_{PRE} = -\frac{1}{N} \sum_{i=1}^N
y_i^u \cdot \log(IP_i^u).
\end{equation}

\noindent The $y_i^u$ in Eq.(37) is the true label of $u$'s interaction behavior to $i$, and the one with the highest predicted probability is compared. Combining the intention loss and sequence loss in the contrastive learning part, the final loss function in the MIDCL training part is the weighted sum of the three losses:

\begin{equation}
\label{eq:MT_3}
\mathcal{L}_{MIDCL} = \mathcal{L}_{PRE} + \alpha \cdot \mathcal{L}_{VAE} + \beta \cdot \mathcal{L}_{SCL}.
\end{equation}

\section{Experiments}\label{sec5}
In this section, we wish to explore the following questions:
                                             
\begin{itemize}
    \item \textbf{RQ1}: How does MIDCL perform compared to other baseline methods?
    \item \textbf{RQ2}: How do the various hyperparameters affect MIDCL?
    \item \textbf{RQ3}: Are modules that intention disentanglement and contrastive learning in MIDCL effective?
    \item \textbf{RQ4}: How is intention optimized during the training of MIDCL?
\end{itemize}

\subsection{Experimental Setting}

\subsubsection{\textbf{Datasets}}
\noindent We choose four real-world datasets, which record the real interaction records of users on different platforms. The specific information about each dataset is shown in Table \ref{tab1}.

\begin{table}[htb]
    \caption{Statistics of the real-world datasets (after preprocessing).}
    \label{tab1}
    \vspace{-5mm}
    \renewcommand{\arraystretch}{1.2}
    \setlength{\tabcolsep}{1mm}{
        \begin{tabular}{cccccc}
            \toprule
            Dataset & \#Users & \#Items & \#Actions & Sparsity & Avg.len\\
            \midrule
            Yelp & 30,431 & 20,033 & 316,354 & 99.95\% & 10.4\\
            Gowalla & 86,168 & 1,271,638 & 6,397,903 & 99.99\% & 74.3\\
            Toys & 19,412 & 11,924 & 167,597 & 99.93\% & 8.6\\
            Beauty & 22,363 & 12,101 & 198,502 & 99.93\% & 8.9\\
            \bottomrule
        \end{tabular}
    }
\end{table}

\begin{itemize}
    \item \textbf{Yelp\footnote{https://www.yelp.com/dataset}}: The dataset is derived from the Yelp ratings network for sentiment computing and social analytics, which contains about 7 million reviews and other business records provided by about 2 million users.
    \item \textbf{Gowalla\footnote{https://snap.stanford.edu/data/loc-gowalla.html}}: This dataset is divided into friend relationship data and location check-in data, and we choose the interaction information of location check-in to model the user's geographic trajectory prediction.
    \item \textbf{Toys \& Beauty\footnote{https://jmcauley.ucsd.edu/data/amazon}}: Contains toys and beauty sales data from Amazon, which can be used for marketing analysis and user purchase behavior prediction.
\end{itemize}

All of the users' records are sorted by timestamp, based on the `5-Core' criterion, all datasets are deleted for each user and item with less than $5$ interactions, and with reference to the leave-one-out strategy, the validation set takes the penultimate item, the test set takes the last item, i.e., the one that has been interacted with most recently, as a positive sample, and then $99$ items from the remaining set of uninteracted items are taken as negative samples, constituting the test set of $1:99$.

In order to enrich the content of the training set, the sequences of the last two items of each cut are sequentially used as the training set labels starting from the second item, and the training portion is all the items before this label, starting from the second item is to ensure that the length of the training portion is at least $1$.

\subsubsection{\textbf{Evaluation Metrics}}
In this experiment, we choose \textit{Hit Ratio} (HR@$K$), \textit{Mean Reciprocal} (MRR) and \textit{Normalized Discounted Cumulative Gain} (NDCG@$K$) to evaluate model performance where $K \in \{5,10\}$.

\subsubsection{\textbf{Baseline Methods}}
We compare MIDCL with the following ten baesline methods.

\noindent (1) \textbf{Machine Learning based methods}.
\begin{itemize}
     \item \textbf{BPR-MF}~\cite{BPR}: Incorporating Matrix Factorization based Collaborative Filtering methods and implicit feedback-based ranking learning methods.
    \item \textbf{FPMC}~\cite{FPMC}: Combining Matrix Factorization with Markov Chain for personalized recommendations.
\end{itemize}

\noindent (2) \textbf{Deep Learning based methods}.
\begin{itemize}
    \item \textbf{Caser}~\cite{Caser}: Mining the local features of sequences using convolution kernels in two different directions, vertically and horizontally, respectively.
    \item \textbf{GRU4Rec}~\cite{GRU4Rec}: Modeling progressive dependencies of user sequence elements via Gated Recurrent Unit (GRU)s.
    \item \textbf{GRU4Rec$^+$}~\cite{GRU+}: Improvements to the loss function, sampling method and regularization technique based on GRU4Rec.
    \item \textbf{SASRec}~\cite{SASRec}: Based on the self-attention mechanism (Transformer) , it can accept longer sequences than GRU and capture potential features among items.
    \item \textbf{BERT4Rec}~\cite{BERT4Rec}: A Transformer-based recommendation model for large sequences, with a pre-training approach and fine-tuning strategy, and with a deep bi-directional self-attention mechanism to capture the dependencies between complex items in user interaction sequences.
\end{itemize}

\noindent (3) \textbf{Contrastive Learning based methods}.
\begin{itemize}
    \item \textbf{S$^3$-Rec${_\text{MIP}}$}~\cite{S3-Rec}: A self-supervised pretraining model based on MIM and proposes four types of learning objectives. In the comparative experiment in this section, we only use MIP i.e. Masked Item Prediction.
    \item \textbf{CL4SRec}~\cite{CL4SRec}: Combining a multi-task contrastive learning model with a multi-head self-attention mechanism, three stochastic enhancement operators are proposed for generating positive and negative samples in contrastive learning.
    \item \textbf{CoSeRec}~\cite{CoSeRec}: A contrastive learning framework based on the Transformer encoder proposes two informative enhancement operators based on CL4SRec, aiming to maintain model robustness.
\end{itemize}

\subsubsection{\textbf{Implementation Details}}
All model parameters in the baselines are taken from the original papers, where fixes the parameters with embedding dimension as 64, batch size as 256 and the maximum sequence length as 50. Our method is implemented in PyTorch, the optimizer is chosen as \textit{Adam}, the learning rate is $1e^{-2}$. The self-attention blocks and attention heads of Transformer encoder as 4, the $\xi$ in $\mathcal{L}_{TRI}$ is 1.5, the $\gamma_c$, $\gamma_m$, $\gamma_r$, $\gamma_i$, and $\gamma_s$ of $AO$ are set to 0.2, 0.3, 0.3, 0.2, 0.3, respectively. The temperature coefficient $\tau$ in $\mathcal{L}_{SCL}$ is taken as 0.8, the weighting factors $\alpha$ and $\beta$ within $\{0.1, 0.2, 0.3, 0.4, 0.5\}$, and the number of intentions ${k}$ within $\{8, 16, 32, 64, 128\}$.

\subsection{Performance Comparison (RQ1)}

\begin{table*}[htb]
\caption{Performance comparisons of different methods. The best score is bolded in each row, and the second best is underlined. The last column shows the relative improvement from our method compared to the best baseline result.}\label{tab2}
\renewcommand{\arraystretch}{1.2}
\resizebox{\textwidth}{!}{
\setlength{\tabcolsep}{1.5mm}{
\begin{tabular}{ll|cc|ccccc|ccc|c|r}
    \toprule%
    \multirow{1}{*}{Dataset} & \multirow{1}{*}{Metric} & \multirow{1}{*}{BPR-MF} & \multirow{1}{*}{FPMC} & 
    \multirow{1}{*}{Caser} & \multirow{1}{*}{GRU4Rec} &
    \multirow{1}{*}{GRU4Rec$^+$} &
    \multirow{1}{*}{SASRec} & \multirow{1}{*}{BERT4Rec} &
    \multirow{1}{*}{S$^3\text{-Rec}_{MIP}$} & \multirow{1}{*} {CL4SRec} & \multirow{1}{*}{CoSeRec}
    & \multirow{1}{*}{MIDCL}
    & Improv. \\
    \hline
    \midrule
    \multirow{5}{*}{Yelp}  
    & HR@5 & 0.4512 & 0.4823 & 0.5123 & 0.5277 & 0.5513 & 0.5672 & \underline{0.5942} & 0.5381 & 0.5726 & 0.5808 & \textbf{0.6375} & 7.29\% \\
    & HR@10 & 0.5877 & 0.6173 & 0.6589 & 0.6854 & 0.7143 & 0.7378 & \underline{0.7683} & 0.7264 & 0.7429 & 0.7548 & \textbf{0.7970} & 3.74\% \\
    & MRR & 0.2908 & 0.3306 & 0.3528 & 0.3659 & 0.3761 & 0.3896 & 0.4142 & 0.3617 & 0.4084 & \underline{0.4167} & \textbf{0.4320} & 3.67\% \\
    & NDCG@5 & 0.3228 & 0.3615 & 0.3547 & 0.3782 & 0.3858 & 0.4098 & \underline{0.4352} & 0.3760 & 0.4130 & 0.4280 & \textbf{0.4669} & 7.28\% \\
    & NDCG@10 & 0.3752 & 0.4128 & 0.4092 & 0.4251 & 0.4476 & 0.4564 & \underline{0.4837} & 0.4355 & 0.4586 & 0.46298 & \textbf{0.5187} & 7.24\% \\
    \midrule
    \multirow{5}{*}{Gowalla}  
    & HR@5 & 0.3260 & 0.3751 & 0.4529 & 0.4397 & 0.4658 & \underline{0.5673} & 0.5624 & 0.4214 & 0.4601 & 0.4586 & \textbf{0.6049} & 6.63\% \\
    & HR@10 & 0.4112 & 0.4581 & 0.5038 & 0.4813 & 0.5160 & \underline{0.6152} & 0.6080 & 0.4697 & 0.5095 & 0.5079 & \textbf{0.6454} & 4.90\% \\
    & MRR & 0.3038 & 0.3494 & 0.3913 & 0.3685 & 0.3962 & 0.5108 & \underline{0.5189} & 0.3515 & 0.3908 & 0.3840 & \textbf{0.5311} & 2.35\% \\
    & NDCG@5 & 0.3134 & 0.3515 & 0.4143 & 0.3782 & 0.4079 & 0.5075 & \underline{0.5186} & 0.3652 & 0.4167 & 0.4012 & \textbf{0.5455} & 5.19\% \\
    & NDCG@10 & 0.3469 & 0.3905 & 0.4431 & 0.4195 & 0.4337 & 0.5230 & \underline{0.5341} & 0.3975 & 0.4418 & 0.4398 & \textbf{0.5587} & 4.61\% \\
    \midrule
    \multirow{5}{*}{Toys}  
    & HR@5 & 0.3074 & 0.3105 & 0.3153 & 0.2935 & 0.3371 & 0.3644 & 0.3530 & 0.3417 & 0.3605 & \underline{0.3758} & \textbf{0.4137} & 10.08\% \\
    & HR@10 & 0.3809 & 0.3862 & 0.3910 & 0.3793 & 0.4176 & 0.4586 & 0.4489 & 0.4273 & 0.4501 & \underline{0.4641} & \textbf{0.5180} & 11.61\% \\
    & MRR & 0.2081 & 0.2152 & 0.2133 & 0.2017 & 0.2329 & \underline{0.2803} & 0.2698 & 0.2414 & 0.2648 & 0.2417 & \textbf{0.2968} & 5.89\% \\
    & NDCG@5 & 0.2226 & 0.2288 & 0.2278 & 0.2174 & 0.2476 & \underline{0.2786} & 0.2675 & 0.2512 & 0.2718 & 0.2609 & \textbf{0.3155} & 13.24\% \\
    & NDCG@10 & 0.2453 & 0.2529 & 0.2651 & 0.2397 & 0.2781 & \underline{0.3089} & 0.2984 & 0.2789 & 0.3157 & 0.3073 & \textbf{0.3491} & 13.01\% \\
    \midrule
    \multirow{5}{*}{Beauty}  
    & HR@5 & 0.3089 & 0.3142 & 0.3145 & 0.3068 & 0.3246 & 0.3719 & 0.3669 & 0.3128 & 0.3970 & \underline{0.4005} & \textbf{0.4339} & 8.34\% \\
    & HR@10 & 0.3968 & 0.4012 & 0.4119 & 0.3879 & 0.4343 & 0.4679 & 0.4641 & 0.4290 & 0.4843 & \underline{0.5044} & \textbf{0.5305} & 5.17\% \\
    & MRR & 0.2274 & 0.2329 & 0.2388 & 0.2388 & 0.2556 & 0.2868 & 0.2837 & 0.2352 & 0.2695 & \underline{0.2940} & \textbf{0.3148} & 7.07\% \\
    & NDCG@5 & 0.2480 & 0.2534 & 0.2548 & 0.2394 & 0.2479 & 0.2854 & 0.2815 & 0.2437 & 0.2871 & \underline{0.3073} & \textbf{0.3347} & 8.92\% \\
    & NDCG@10 & 0.2534 & 0.2714 & 0.2681 & 0.2499 & 0.2700 & 0.3063 & 0.3058 & 0.2686 & 0.3003 & \underline{0.3253} & \textbf{0.3660} & 12.51\% \\
    \bottomrule
\end{tabular}}}
\end{table*}

Table \ref{tab2} shows the performance comparisons results between several baseline methods and MIDCL. Based on the table, the following conclusions can be summarized. First, BPR-MF and FPMC perform poorly compared to other deep learning methods, and contrastive learning improves model performance in some scenarios. For example, in \textit{Beauty}, CoSeRec performs better than other methods except MIDCL, but the Transformer-based methods such as SASRec and BERT4Rec perform better in \textit{Gowalla}.

Moreover, The performance of S$^3$-Rec${_\text{MIP}}$ is the worst among the 3 models based on contrast learning, mainly because only the former MIP self-supervised task was used for training in the experiments, whereas both CL4SRec and CoSeRec augmented the user's sequences and were able to achieve better results in the process of contrastive learning; CoSeRec was supplemented with two additional augmentation factors in comparison to CL4SRec, and thus performs better under most of the datasets.

Finally, MIDCL outperforms other methods on all four real datasets because it combines intention disentanglement and contrastive learning on top of the Transformer encoder, enabling it to better understand users' interactive intentions. On the HR@10 metric, MIDCL leads the next best performing model by 2.87\%, 3.02\%, 5.39\%, and 2.61\% in the four datasets, respectively.

\subsection{Hyper-parameter Sensitivity Analysis (RQ2)}
In this section, we verify the hyper-parameter sensitivity of MIDCL with respect to the weighting factors $\alpha \& \beta$ and the number of intentions ${k}$ in the four datasets mentioned above, respectively.

\begin{figure}[htb]
    \centering
    \subfigure[]{
        \includegraphics[width=0.46\columnwidth]{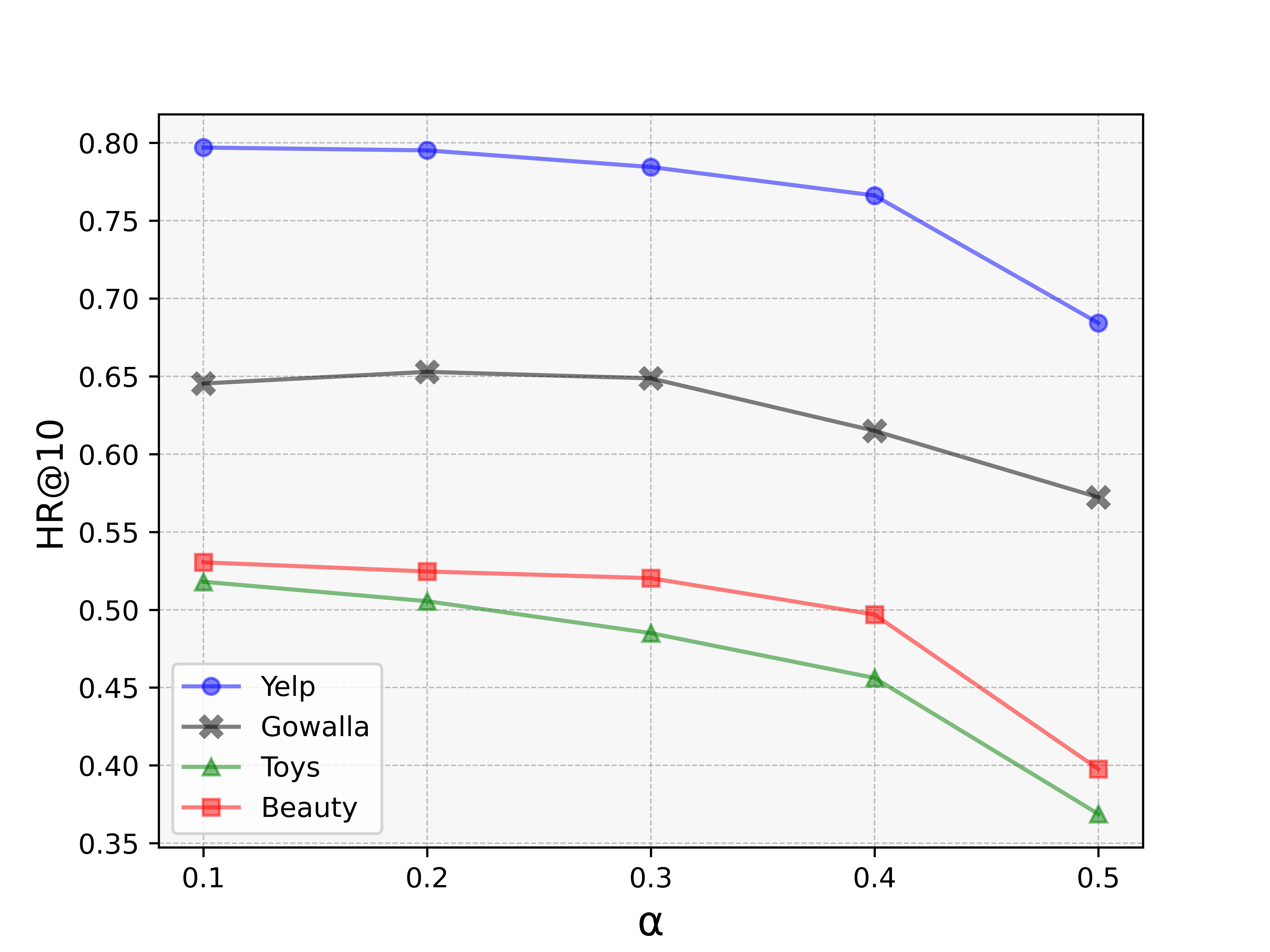}\ 
    }
    \subfigure[]{
        \includegraphics[width=0.46\columnwidth]{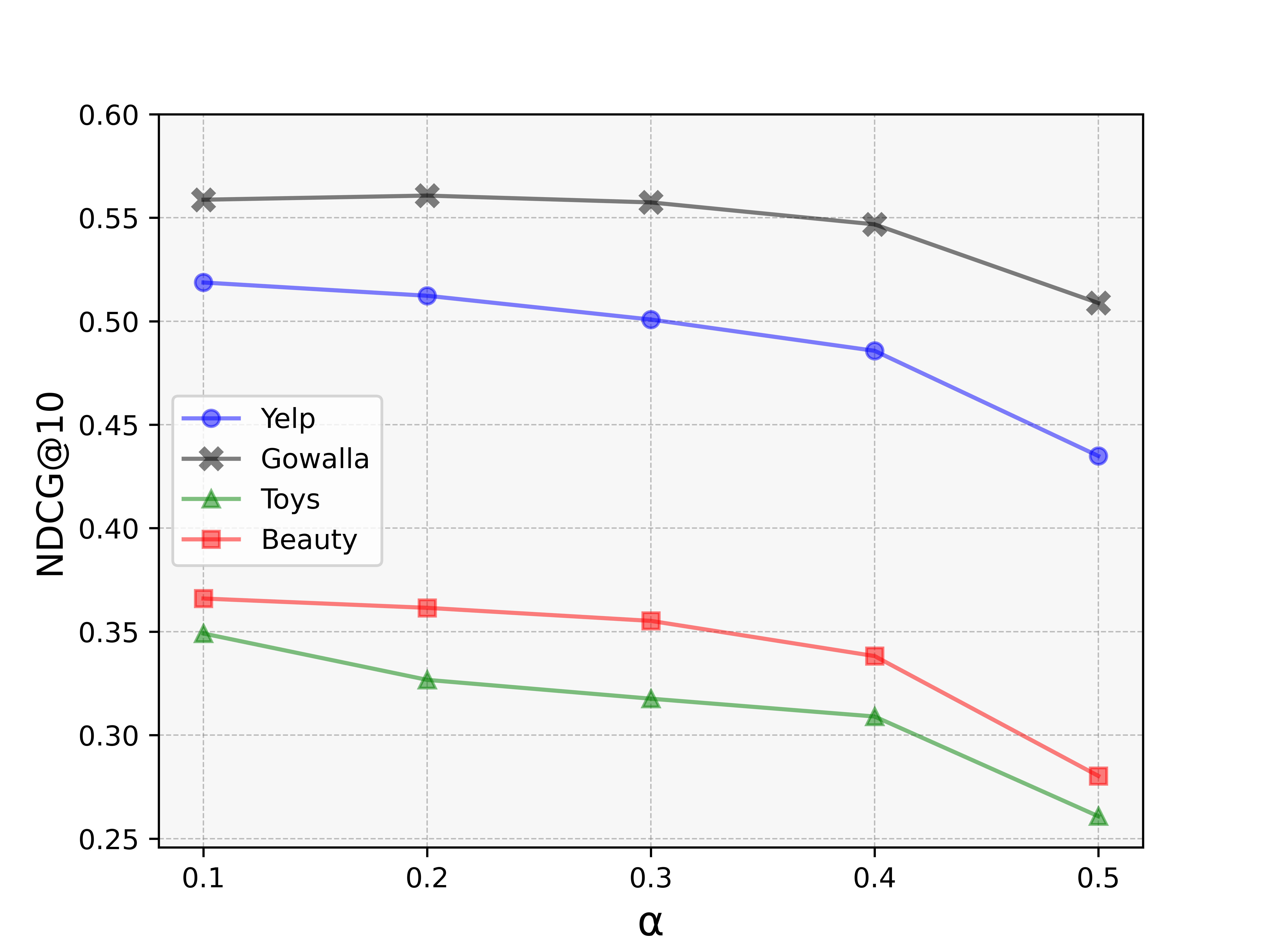}
    }
    \caption{Performance comparison w.r.t. Hyper-Parameter $\alpha$.}
    \label{fig3}
\end{figure}

Fig.\ref{fig3} depicts the effect of the Hyper-Parameter $\alpha$ on the performance, where fixing $\beta$ as 0.3 and ${k}$ as 8. we can see that each dataset essentially reaches its best performance at $\alpha = 0.1$. Because the loss function of VAE consists of the KLD of the latent spatial prior distribution and the ternary loss, if the weighting factor is increased, the decline of the main loss, i.e. the cross-entropy loss function, is gradually affected. Especially when $\alpha$ reaches 0.5, the metrics of each dataset basically fall off a cliff.

\begin{figure}[htb]
    \centering
    \subfigure[]{
        \includegraphics[width=0.46\columnwidth]{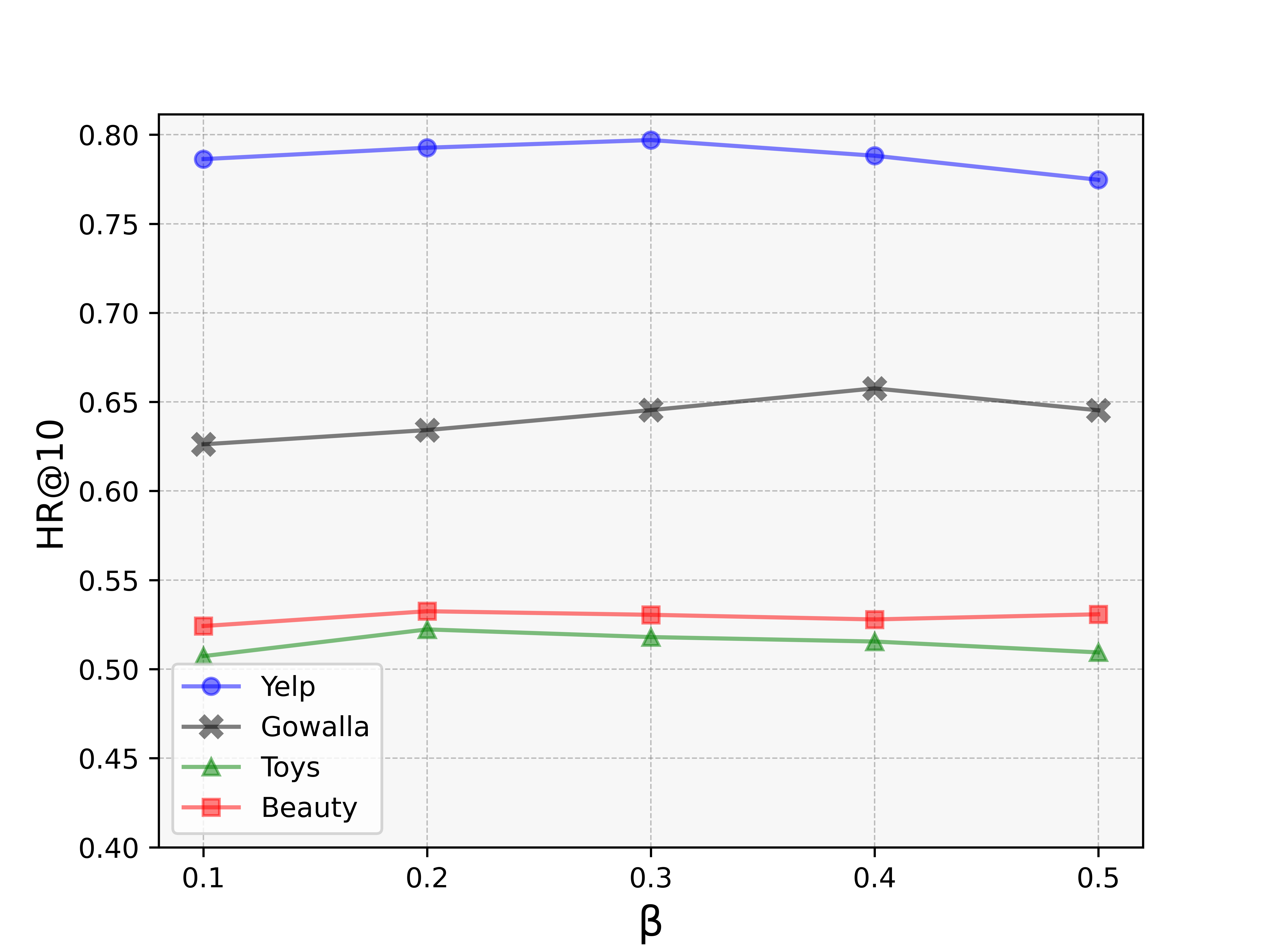}\ 
    }
    \subfigure[]{
        \includegraphics[width=0.46\columnwidth]{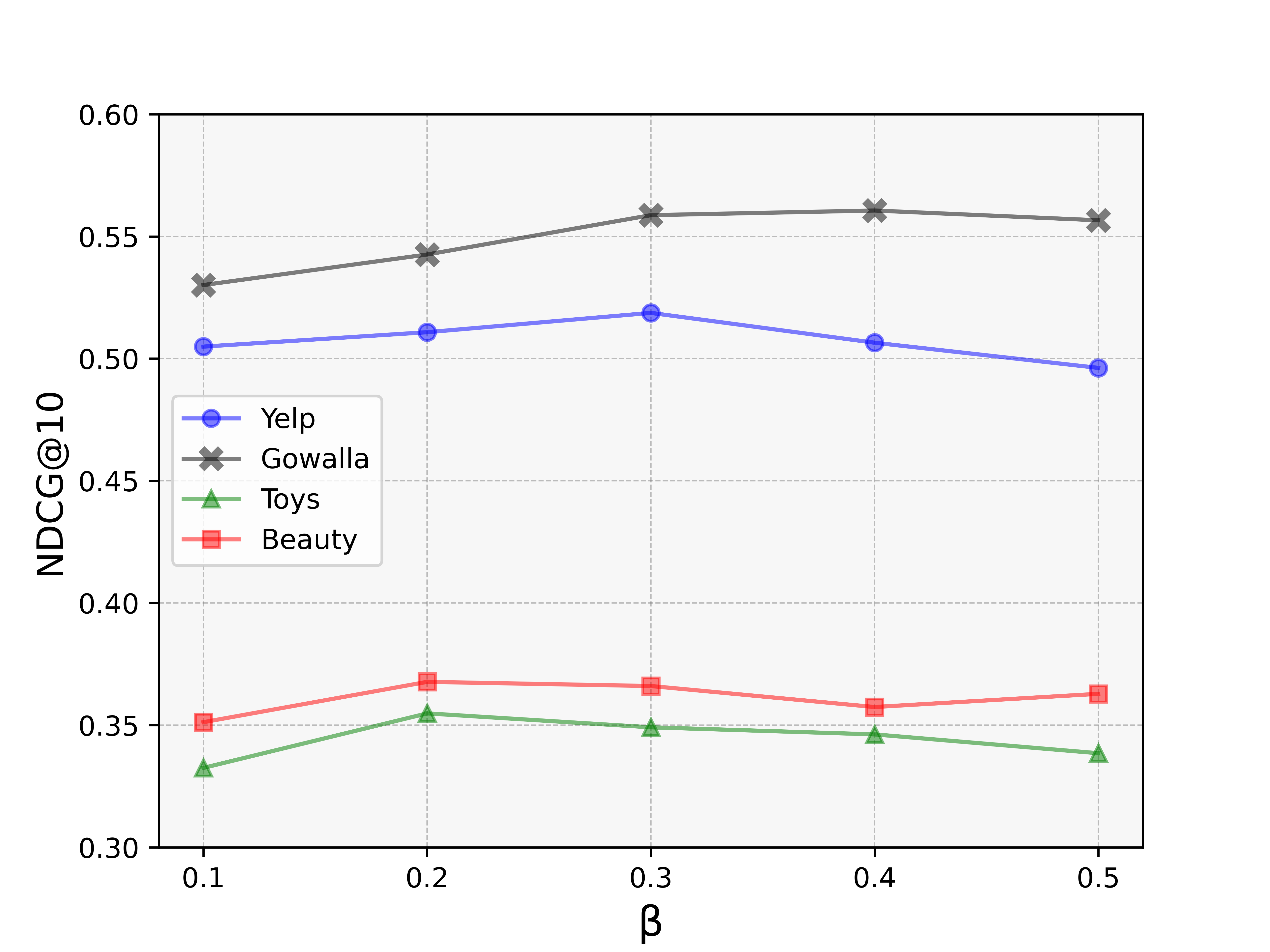}
    }
    \caption{Performance comparison w.r.t. Hyper-Parameter $\beta$.}
    \label{fig4}
\end{figure}

Fig.\ref{fig4} shows the effect of the Hyper-Parameter $\beta$ on the performance, where fixing $\alpha$ as 0.1 and ${k}$ as 8. As shown, the best-performing phases of each dataset are basically in the range of $\left[0.2, 0.4\right]$ for $\beta$, with \textit{Gowalla} having the best performance at $\beta=0.4$, and \textit{Toys} \& \textit{Beauty} having the best at $\beta=0.2$. In particular, the metrics and trends in \textit{Toys} and \textit{Beauty} are close to each other because they are both part of the same Amazon dataset, but \textit{Gowalla} is a different type of large dataset.

\begin{figure}[htb]
    \centering
    \subfigure[]{
        \includegraphics[width=0.46\columnwidth]{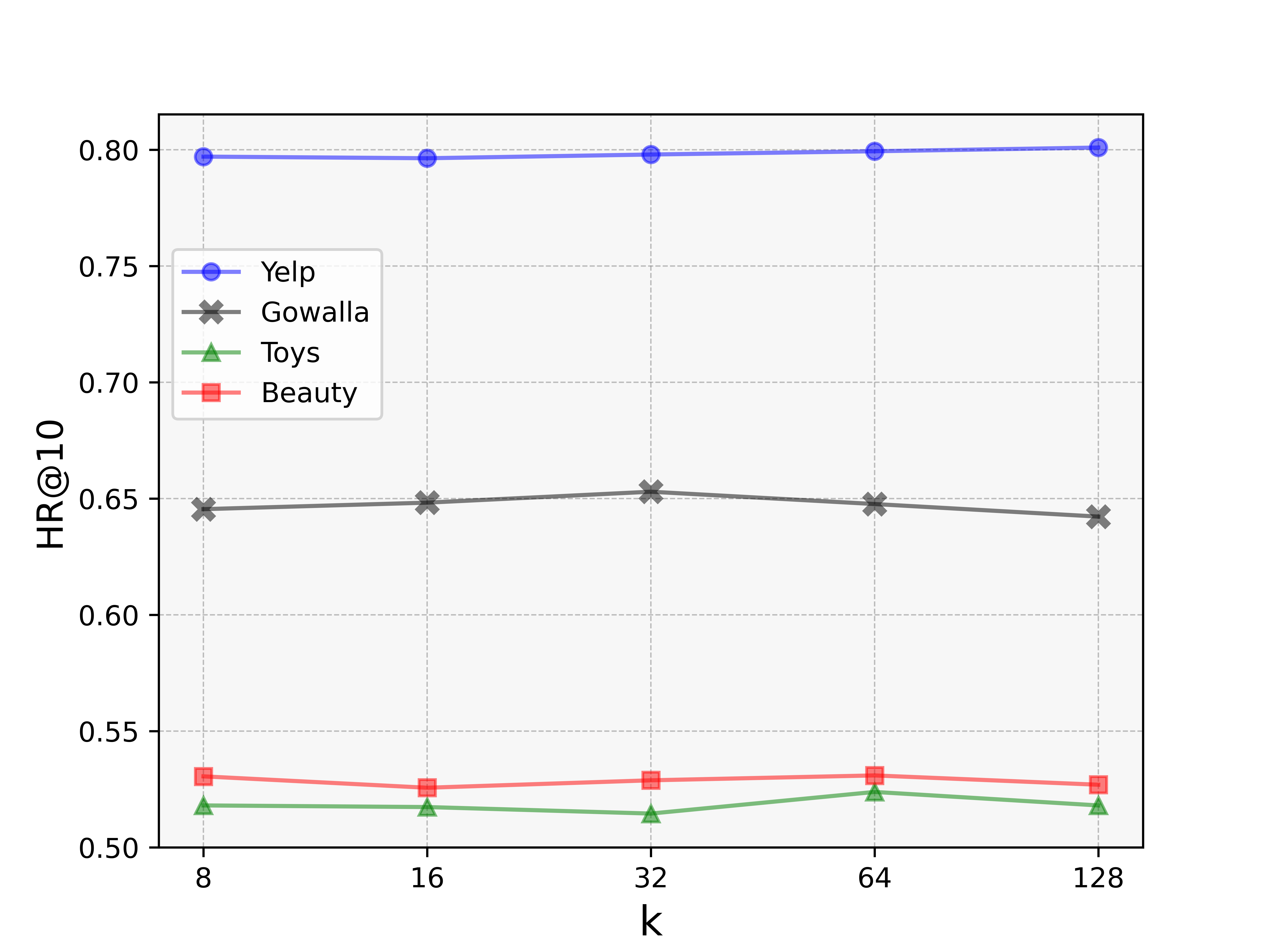}\ 
    }
    \subfigure[]{
        \includegraphics[width=0.46\columnwidth]{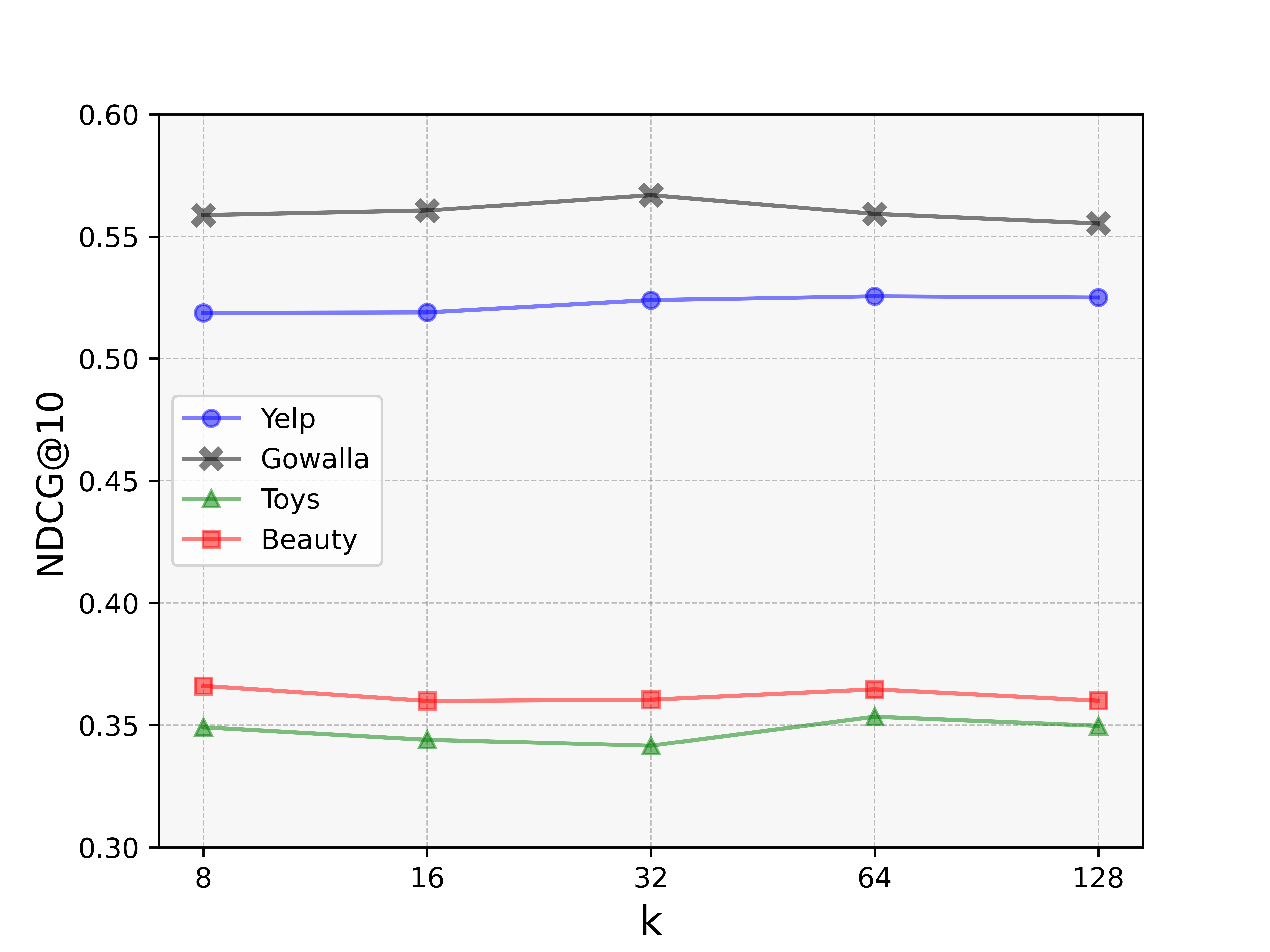}
    }
    \caption{Performance comparison w.r.t. the number of intentions ${k}$ where $\alpha$ is 0.1 and $\beta$ is 0.3.}
    \label{fig5}
\end{figure}

Fig.\ref{fig5} describes the influence of number of intentions ${k}$. We can find that none of the datasets are sensitive to ${k}$, with \textit{Toys} showing a slight decrease in performance at $k=32$, and \textit{Gowalla}, on the contrary, showing the best performance at this point. If the change of this hyper-parameter does not bring a large change to the model during the training process, the speed of convergence as well as the resource consumption will serve as a valuable metric for reference, so the choice of $k=8$ will not bring an impact on the performance of the model as well as significantly reduce the time of training and loss convergence.

\subsection{Ablation Study (RQ3)}

To verify the validity of each module of MIDCL, the intention disentanglement module of VAE, the contrastive learning module, and the original encoder module only for sequential recommendation are removed from the model, and the experiments are conducted under the conditions of $k=8$, $\alpha=0.1$ and $\beta=0.3$, respectively.

\begin{table}[htb]
  \caption{Ablation study of MIDCL on \textit{Yelp} and \textit{Beauty} (H means HR and N means NDCG).}
  \label{tab3}
  \renewcommand{\arraystretch}{1.2}
  \setlength{\tabcolsep}{1mm}{
    \begin{tabular}{ll|cccc}
        \toprule
        Dataset & Model & H@5 & H@10 & N@5 & N@10\\
        \midrule
        \multirow{4}{*}{Yelp}
        & MIDCL & \textbf{0.6375} & \textbf{0.7970} & \textbf{0.4669} & \textbf{0.5187}\\
        & w/o MID & 0.6186 & 0.7691 & 0.4584 & 0.5073\\
        & w/o CL & 0.6164 & 0.7736 & 0.4441 & 0.4951\\
        & only SR & 0.5914 & 0.7566 & 0.4288 & 0.4822\\
        \midrule
        \multirow{4}{*}{Beauty}
        & MIDCL & \textbf{0.4339} & \textbf{0.5305} & \textbf{0.3347} & \textbf{0.3660}\\
        & w/o MID & 0.4069 & 0.5040 & 0.3129 & 0.3442\\
        & w/o CL & 0.3905 & 0.4804 & 0.3098 & 0.3325\\
        & only SR & 0.3864 & 0.4764 & 0.3022 & 0.3312\\
        \bottomrule
    \end{tabular}}
\end{table}

Table \ref{tab3} records the results of each ablation model on the evaluation metrics. As shown above, both the intention disentanglement and contrastive learning modules contribute to MIDCL, and both modules are more effective than simple embedding encoder module for sequential recommendation only, with the importance of contrastive learning being more pronounced.

\subsection{Intention Visualization (RQ4)}
We design the visualization process of the intention and user embedding to verify our method and figure out how the disentangled intention changes during the training process. In order to present the results on a two-dimensional plane, we choose t-SNE to reduce feature dimensionality and combine it with K-Means clustering to observe the distance between intentions and user's features.

\begin{figure*}[htb]
    \centering
    \subfigure[]{
        \includegraphics[width=0.29\linewidth]{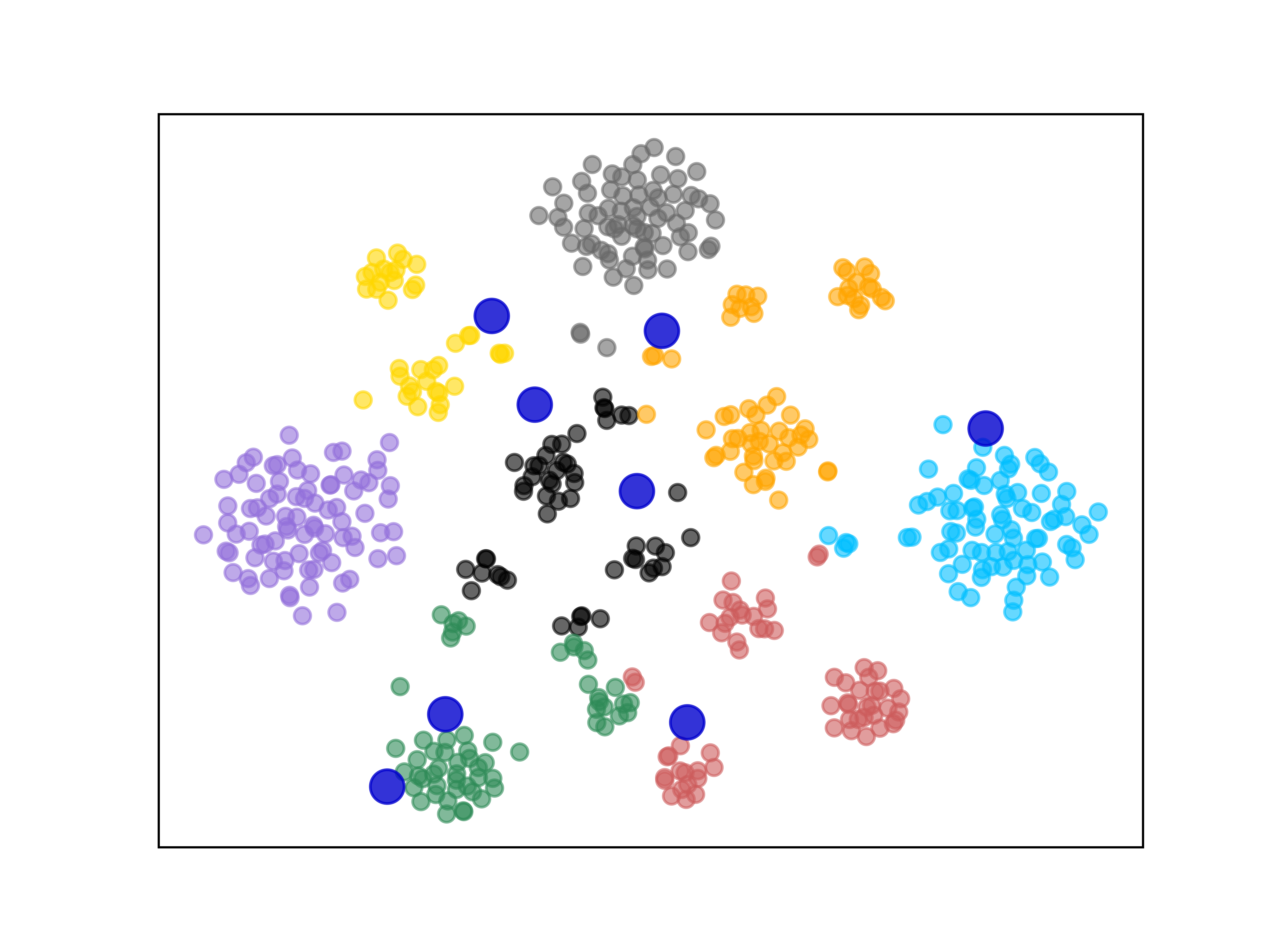}\ 
    }
    \subfigure[]{
        \includegraphics[width=0.29\linewidth]{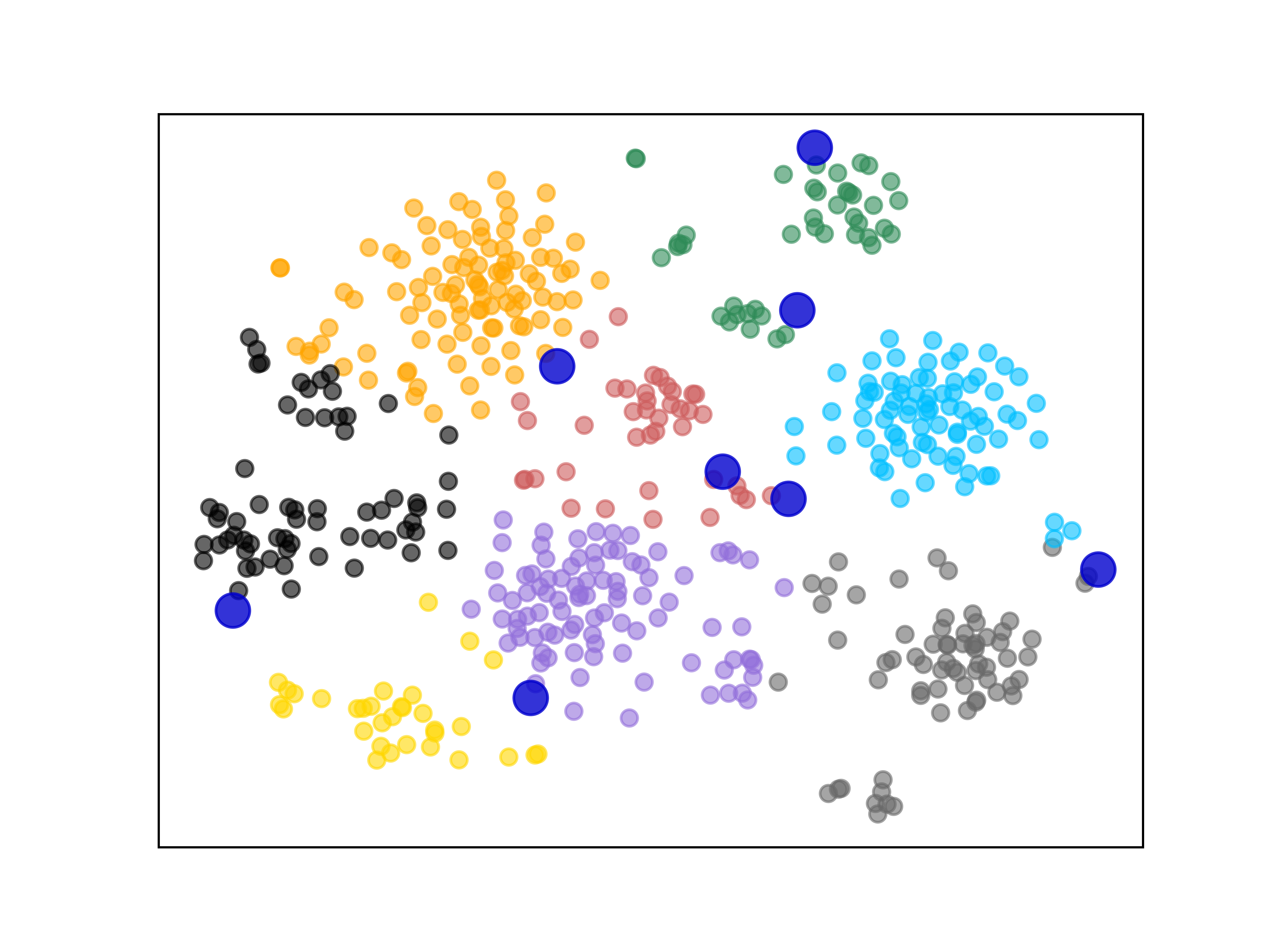}
    }
    \subfigure[]{
        \includegraphics[width=0.29\linewidth]{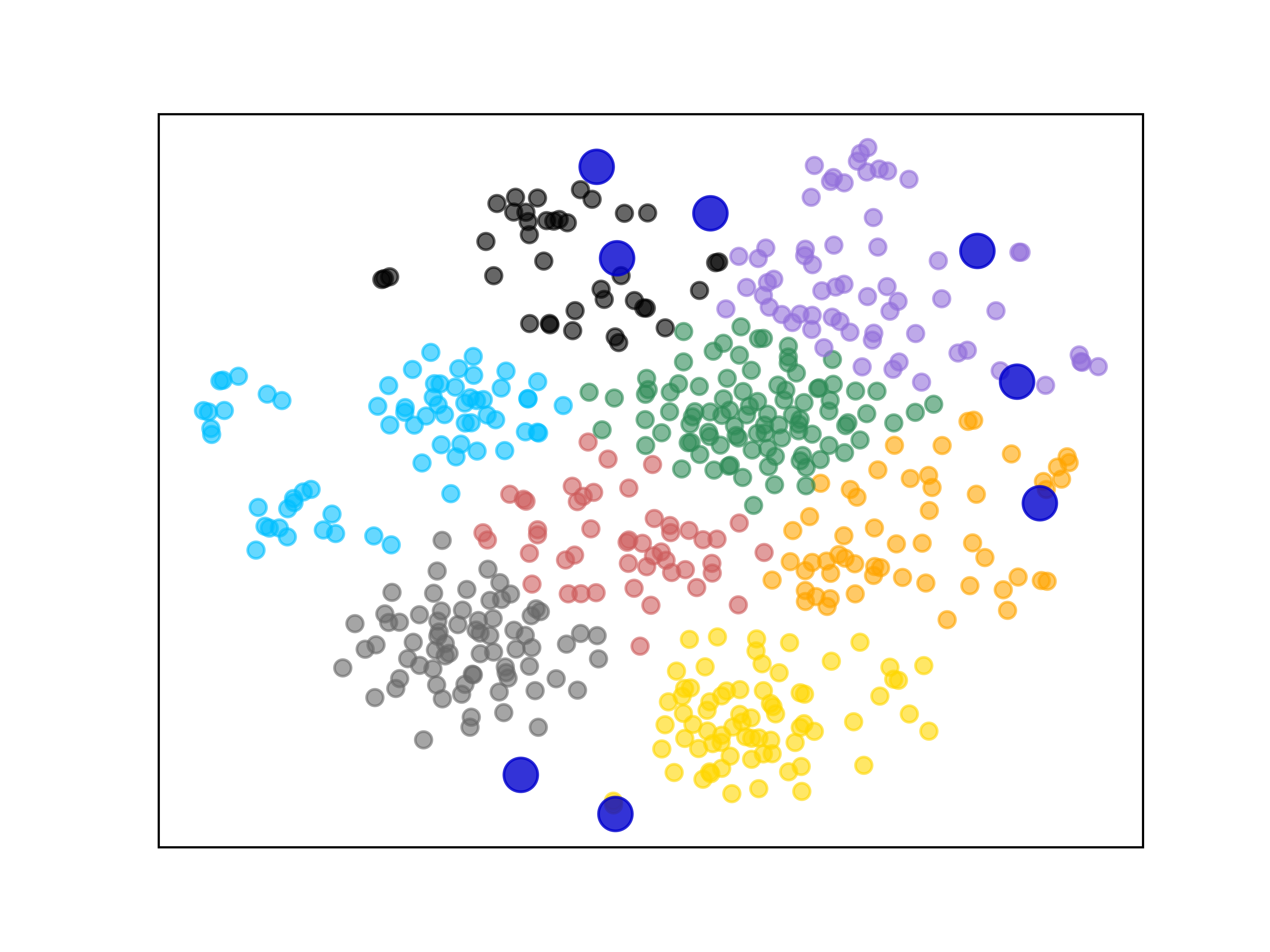}
    }
    \subfigure[]{
        \includegraphics[width=0.29\linewidth]{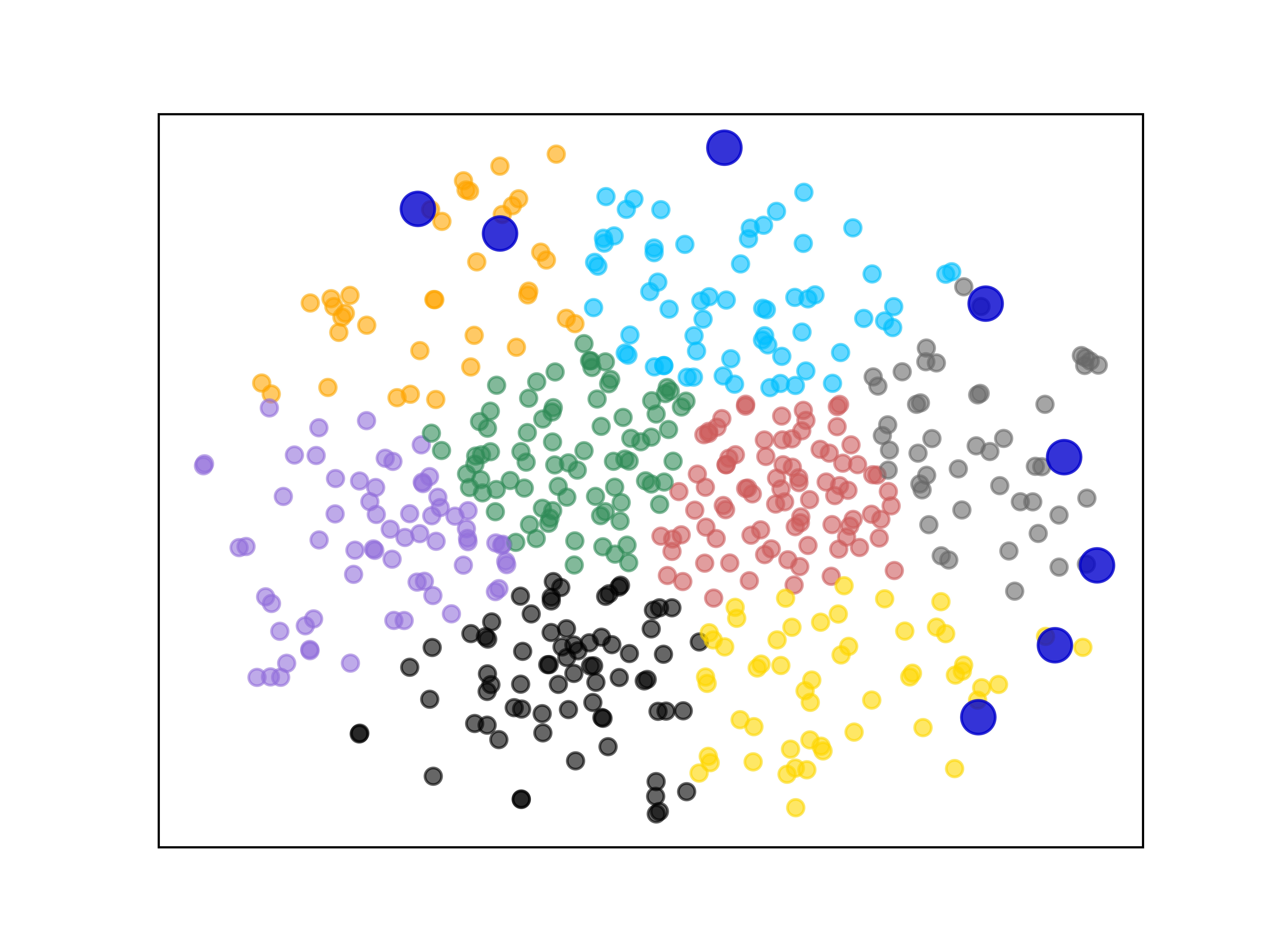}
    }
    \subfigure[]{
        \includegraphics[width=0.29\linewidth]{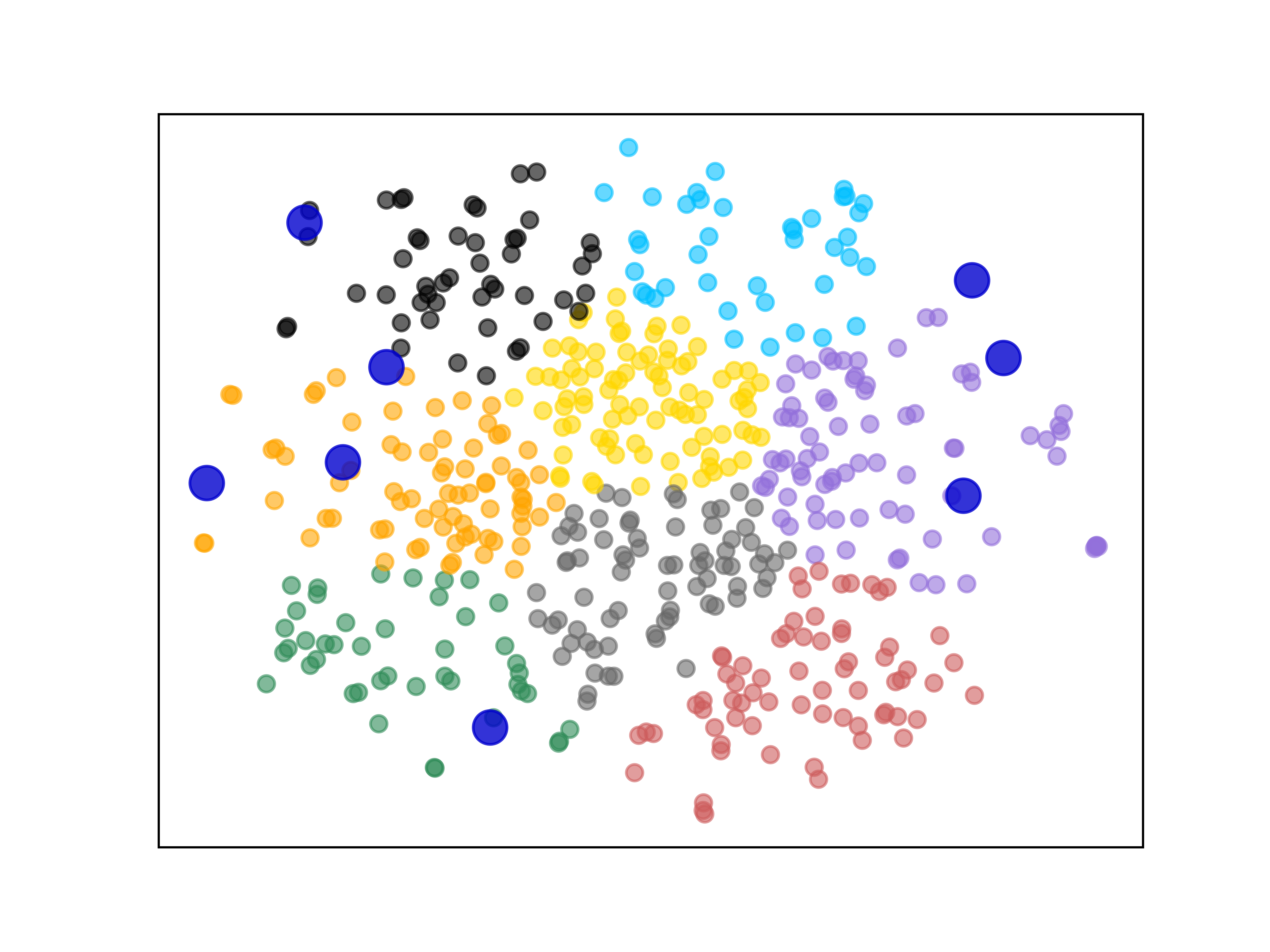}
    }
    \subfigure[]{
        \includegraphics[width=0.29\linewidth]{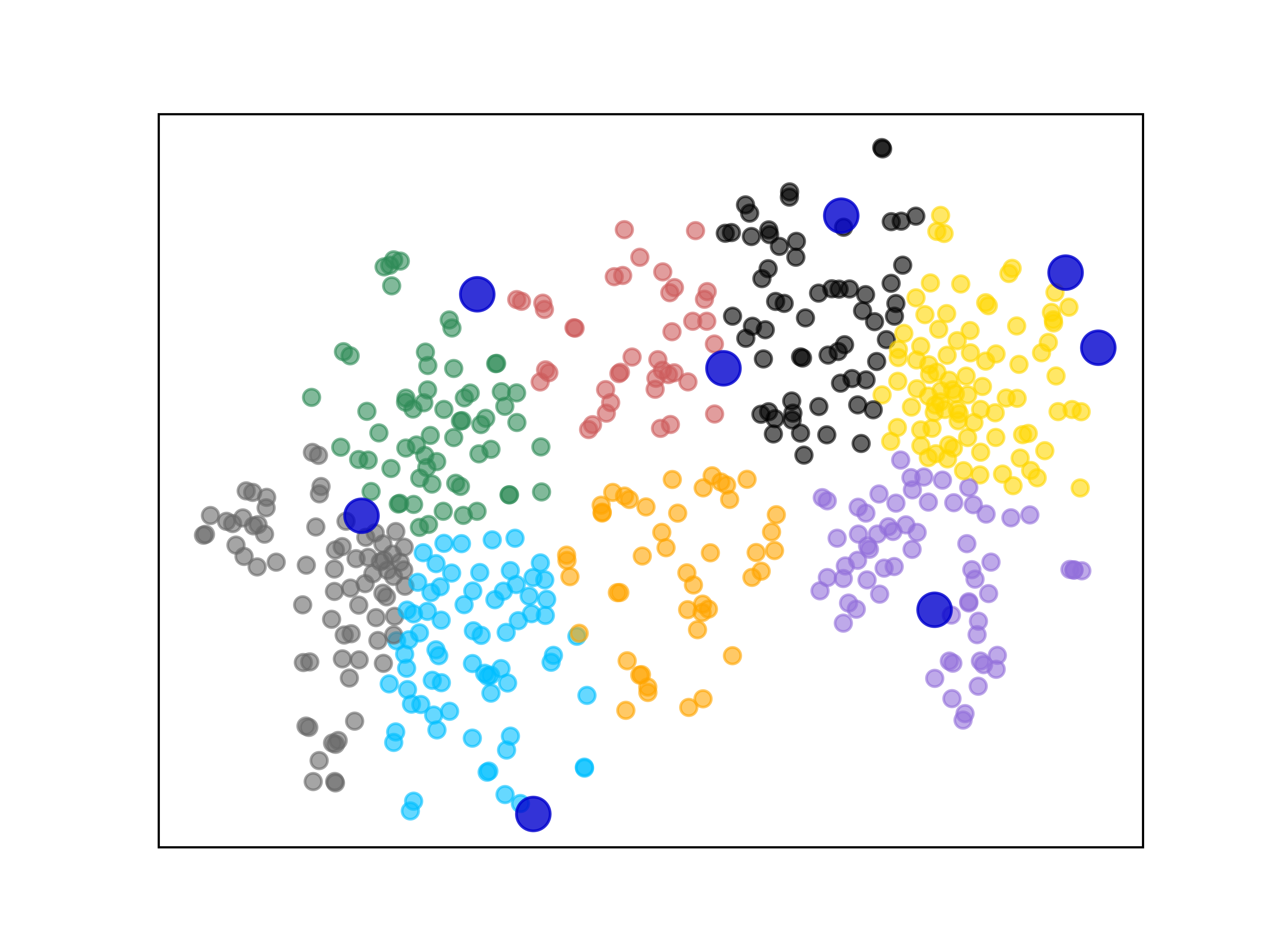}\ 
    }
    \subfigure[]{
        \includegraphics[width=0.29\linewidth]{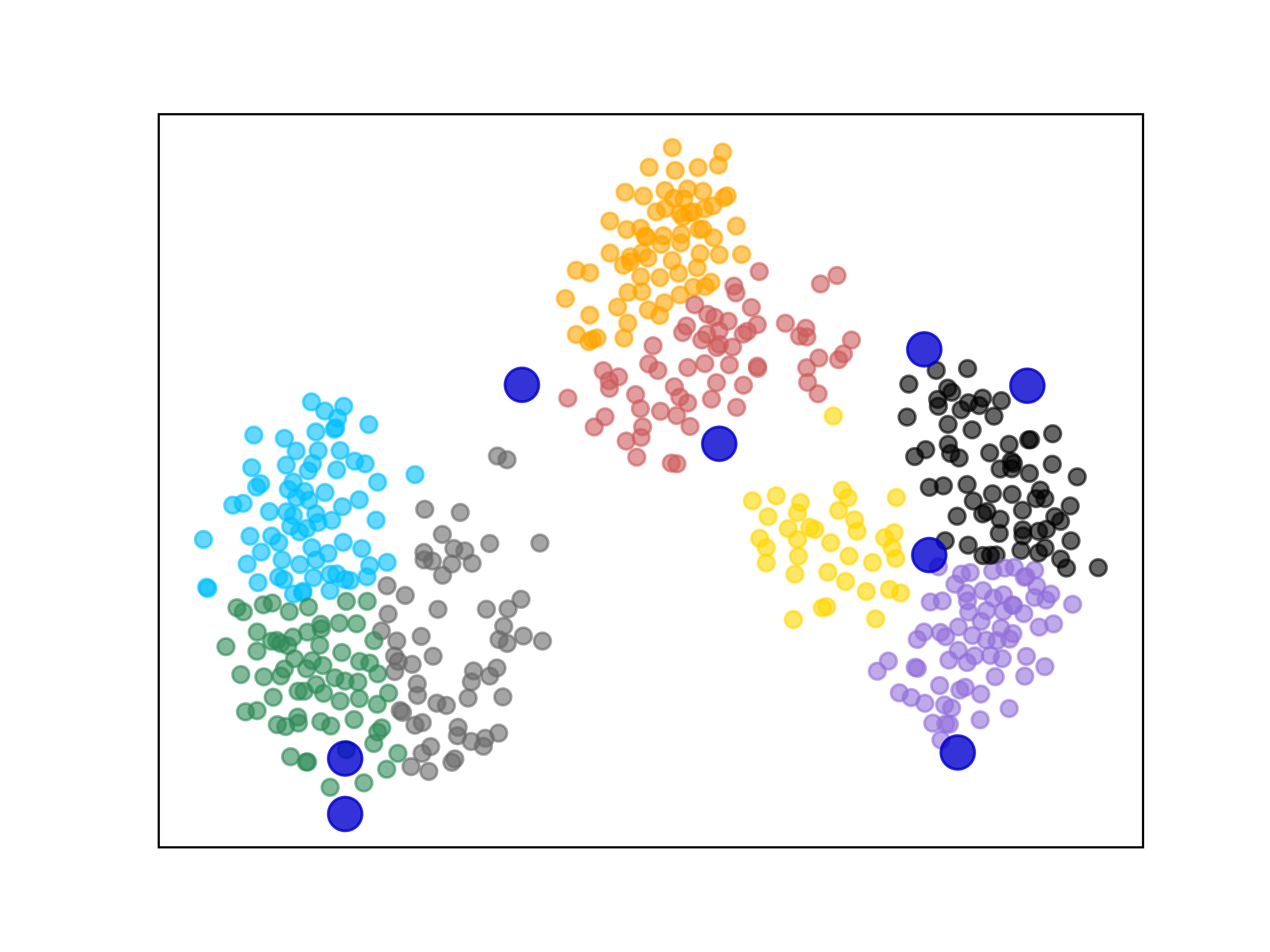}
    }
    \subfigure[]{
        \includegraphics[width=0.29\linewidth]{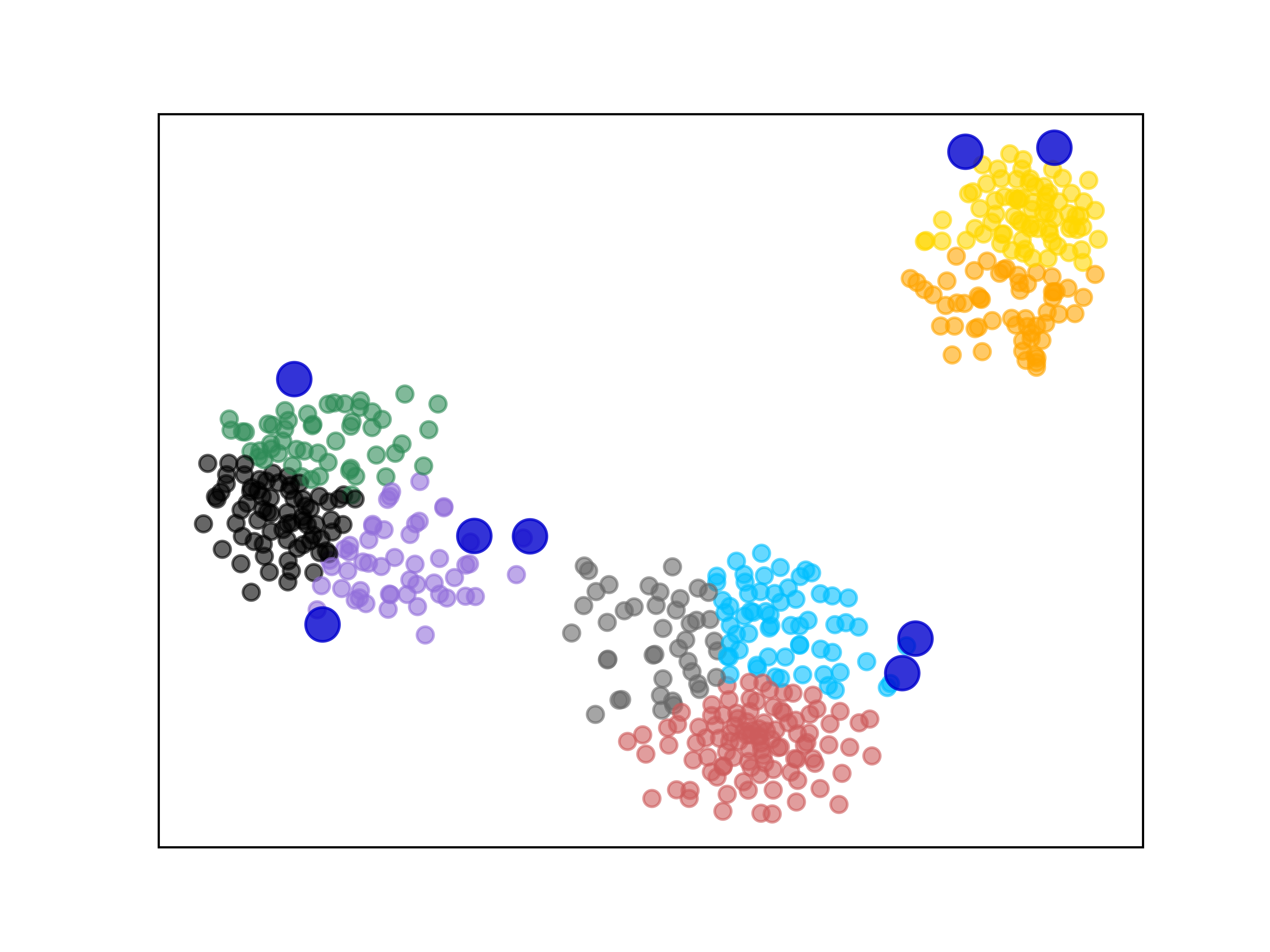}
    }
    \subfigure[]{
        \includegraphics[width=0.29\linewidth]{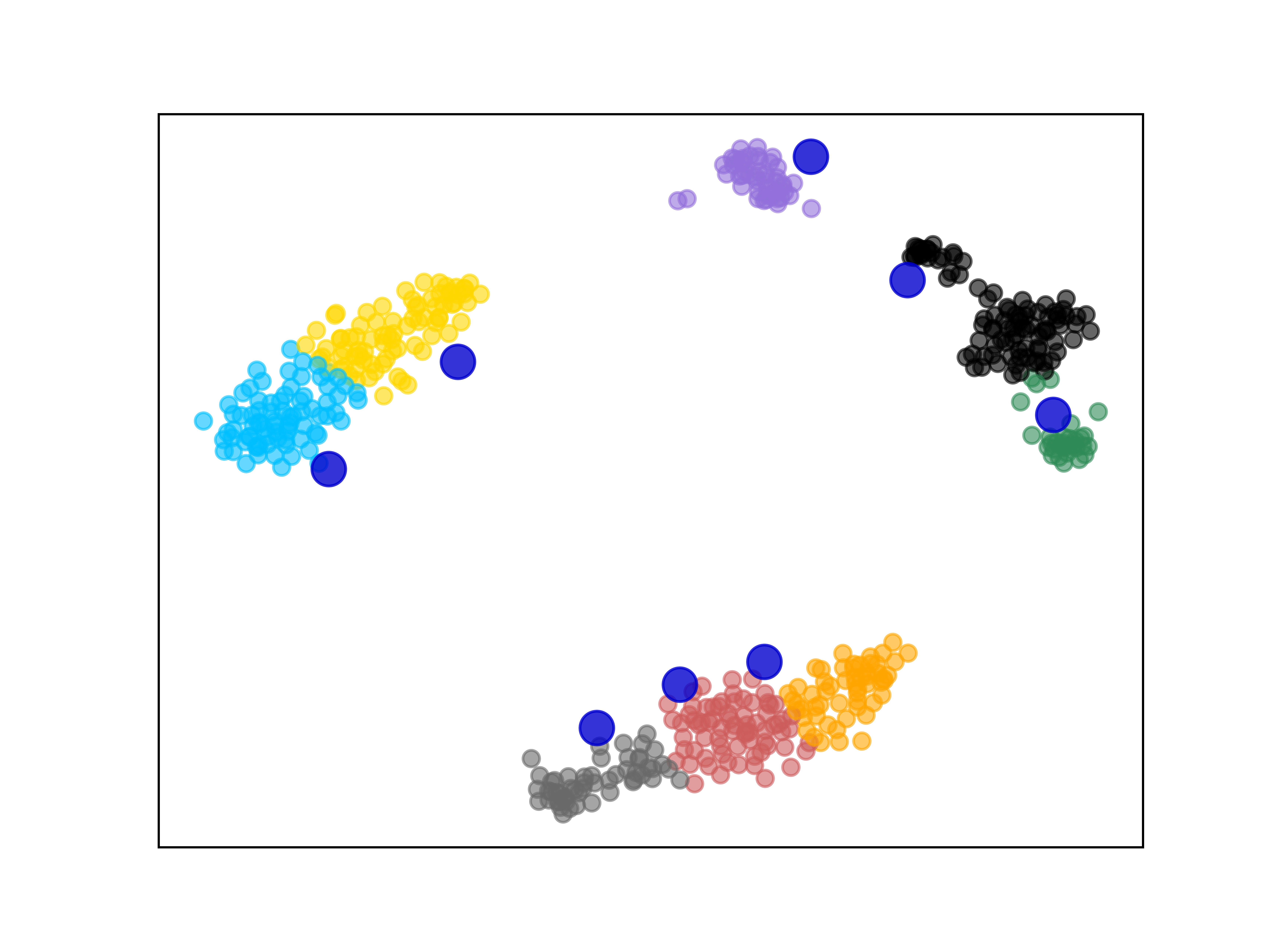}
    }
    \caption{2D visualization of intention and user features on \textit{Yelp}.}
    \label{fig6}
\end{figure*}

Fig.\ref{fig6} illustrates the different stages of a certain training process in which the sequence and intention features of users are clustered after dimensionality reduction to the two-dimensional plane, where the large dark blue dots indicate the disentangled $k$-intents, and this time $k=8$ is used as an example. The remaining small dots of eight different colors indicate the sequence features of the users in one batch-size, the same color represents the same clusters, and the features of the users with the same clusters means they have a higher degree of similarity.

As shown in the figure, (a) represents the beginning of the training of all the user's sequence features present a number of clusters of different sizes and randomly distributed; subsequently, from (b) to (e), the user features began to gradually spread; as the gradient of the loss decreases, each user began to draw closer to its similar intentions from (f) to (h); at the end in (i), when the iteration tends to stabilize, all users basically have found the closest to their own intention representation, and constitute the $k$ clusters. With given data, it is reasonable to assume that users within the same cluster have the most similar intention, and they are more likely to produce similar even same interaction behaviors in the next moment.

\section{Conclusion}\label{sec6}
In this paper, we propose a contrastive learning method MIDCL based on multi-intention disentanglement for sequential recommendation, which enables the prediction of the next item in a sequence by mining user's intention. First, the behavioral sequences are embedded by Transformer encoder; second, multi-intentions of users are disentangled based on VAE in the potential space with a parameter-sharing Transformer encoder, whose dimensions are used as the number of user's latent intentions; finally, we propose two contrastive learning paradigms to optimize the feature representations of the intentions and sequences, respectively, to weaken the negative impact of irrelevant intentions. The experimental results on four real-world datasets show that MIDCL achieves the optimal results in all baseline methods and validates the contribution of the intention disentanglement and contrastive learning modules. To explore changes in intention with training, we also conduct the visualization of the intentions as a way to demonstrate the feasibility and interpretability of intention disentanglement module.

For future work, we will focus on evaluating the incremental impact of each user action on intentionality, as well as the disentanglement and analysis of multiple intentions during each interaction, and consider the contextual information from the interaction scenario and the social network, etc., which will yield a more comprehensive and personalized representation of the user's interactive intention.

\bmhead{Acknowledgements}

This paper is supported by the National Key Research and Development Program of China (314) and the Independent Project of National Key Laboratory (2024-SKL-005).

\end{document}